# CIGaRS I: combined simulation-based inference from type Ia supernovae and host photometry

Konstantin Karchev [1,2], Roberto Trotta [1,3,4,5] & Raúl Jiménez [2,6]

Using type Ia supernovae as cosmological probes requires empirical corrections that are correlated with their host environment. Here we present a unified Bayesian hierarchical model designed to infer, from purely photometric observations, the intrinsic dependence of the brightness of type Ia supernovae on progenitor properties (metallicity and age), the delay-time distribution that governs their rate as a function of age and cosmology, as well as the redshifts of all hosts. The model incorporates physics-based prescriptions for star formation and chemical evolution from Prospector-β, dust extinction of both galaxy and supernova light, and observational selection effects. We show with simulations that intrinsic dependences on metallicity and age have distinct observational signatures, with metallicity mimicking the well-known step of magnitudes of type Ia supernovae across a host stellar mass of ~$10^{10}$ $M_{\odot}$. We then demonstrate neural-simulation-based inference of all model parameters from mock observations of ~16,000 type Ia supernovae and their hosts up to redshift 0.9. Our joint physics-based approach delivers robust and precise photometric redshifts (~0.01 median scatter) and improves cosmological constraints by a factor of ~4 over analyses of the small fraction of objects with spectroscopic follow-up. This approach unlocks the full power of photometric data and paves the way for an end-to-end simulation-based analysis pipeline in the LSST era.

It has long been known that the brightnesses, stretches and colours of type Ia supernovae are correlated with the global properties of their host galaxies, for example morphology, stellar mass, star-formation rate (SFR), metallicity and stellar age[1–12]. Similar dependences on the local environment of the supernovae have also been found[13–27]. In current cosmological analyses (for example, refs. 28–30), which use type Ia supernovae as standardizable candles, it is common practice to (attempt to) correct for all these effects en masse via an ad hoc 'mass step': a magnitude offset between type Ia supernovae hosted in galaxies with stellar masses above and below a given threshold, usually set to the sample median or fixed around $10^{10}$ $M_{\odot}$. The rationale is that stellar mass is a simple-to-estimate proxy for nearly all host properties. Unveiling the physical processes underlying host–type Ia supernova connections would, thus, not only enhance our understanding of type Ia supernovae and their formation but also deliver more accurate and precise constraints on dark energy by improving the standardization procedure.

[1]Theoretical and Scientific Data Science, International School for Advanced Studies (SISSA), Trieste, Italy. [2]Institute of Cosmos Sciences (ICC), University of Barcelona, Barcelona, Spain. [3]Department of Physics, Imperial College London, London, UK. [4]Institute for Fundamental Physics of the Universe (IFPU), Trieste, Italy. [5]Italian Research Center on High-Performance Computing, Big Data and Quantum Computing (ICSC), Casalecchio di Reno, Italy. [6]Catalan Institution for Research and Advanced Studies (ICREA), Barcelona, Spain. e-mail: kkarchev@sissa.it

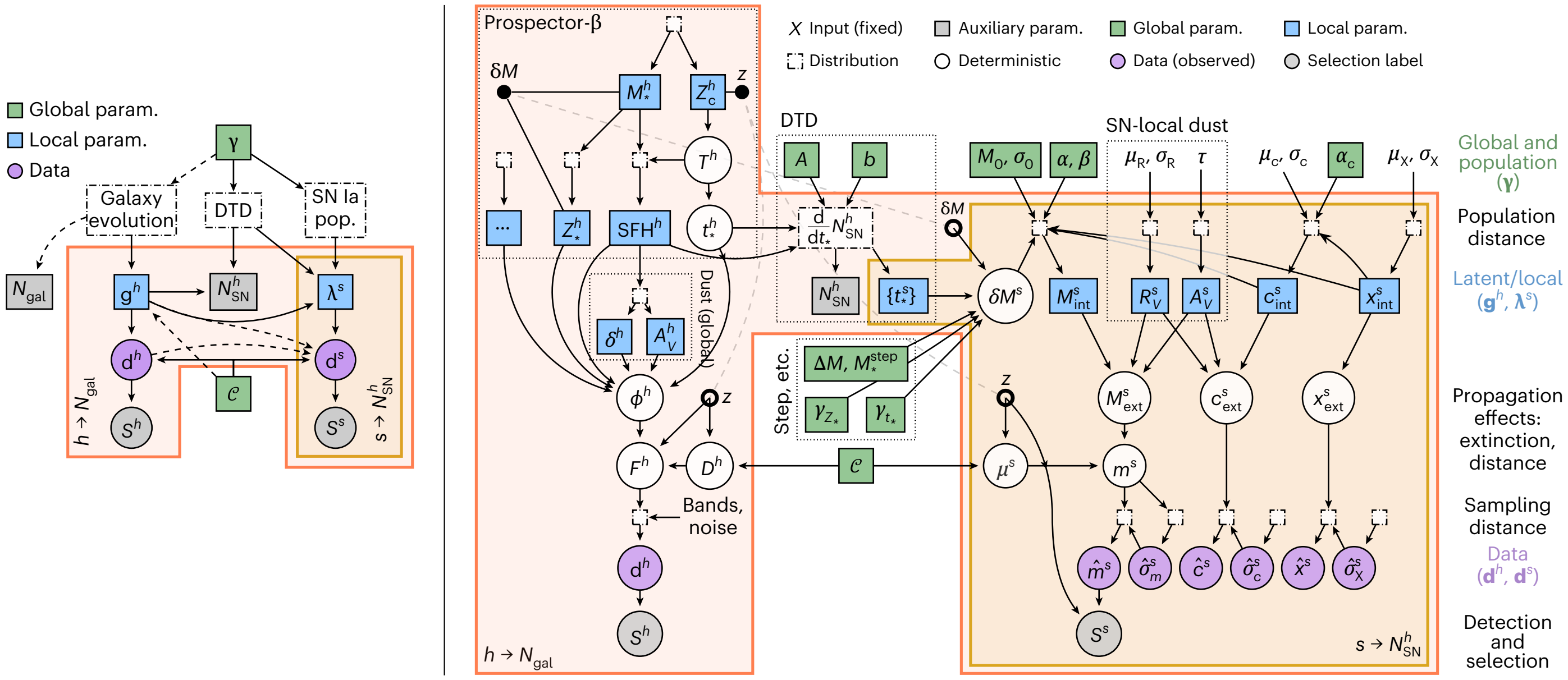

**Fig. 1 | Unified Bayesian hierarchical modelling of type Ia supernovae and host galaxies.** Left: high-level overview. Right: our simulator in detail. It combines Prospector-β[59,60] and Simple-BayeSN[32] via a 'bridge' formed by the DTD and intrinsic host–supernova correlations to produce self-consistent LC summaries for type Ia supernova and galaxy photometry (subject to sample selection). Symbol definitions and priors are listed in Extended Data Table 1, with full details given in section 'Unified forward modelling of galaxies and type Ia supernovae'. param., parameter; SN, supernova; SN Ia pop., type Ia supernova population; dist., distribution.

The environment around a supernova also has extrinsic effects on its light, which is dimmed and reddened by dust along the line of sight, both in the immediate surroundings of the progenitor system and in the interstellar space within the host. It can also be scattered into the line of sight, which noticeably affects finite-resolution photometry[31]. Accounting for dust requires careful hierarchical modelling[32], as (1) extinction has a similar effect (redder is dimmer) to intrinsic colour-related standardization (bluer is brighter) and (2) the amount and properties of the dust in a galaxy are related to its SFR and stellar mass (for example, refs. 33,34) and, hence, to the ages and metallicities of the progenitors of the type Ia supernovae. Different studies have come to conflicting conclusions regarding this interplay. Brout and Scolnic[35] attributed the mass step entirely to differences in the dust in low- and high-mass hosts rather than any intrinsic, physical or causal effect, whereas the analysis of Thorp and Mandel[36] prefers a single dust law for all hosts and a residual mass step of ~0.05 mag, with more massive galaxies hosting intrinsically brighter type Ia supernovae. By contrast, a unified Bayesian model comparison using the same data and models[37] disfavoured a non-zero mass step (posterior odds of 2:1), whereas Grayling et al.[38] and Grayling and Popovic[39] found evidence in more recent datasets both for intrinsic differences in the brightness and stretch distributions of type Ia supernovae as well as different dust properties (wavelength dependence and optical depths) in low- and high-mass hosts. Moreover, Wiseman et al.[40,41] argued in favour of a split in host age rather than mass, considering that the two are connected by galactic evolution through the process known as 'downsizing' (see, for example, ref. 42).

Resolving these degeneracies and arriving at a definite conclusion regarding host–type Ia supernova connections require tackling all interrelated effects simultaneously, including the apparent dimming due to cosmological distance and selection biases. Although Bayesian hierarchical modelling provides the principled framework for doing this, traditional Markov chain Monte Carlo methods require a joint sampling of the vast number of latent parameters (tens per supernova) and explicit computation—at every step—of the selection probability, which is intractable in realistic scenarios. To date, analyses either resort to ad hoc approximations[29,43,44] or are split in two separate steps (see, for example, refs. 36,45): first, the distance moduli of supernovae (that is intrinsic brightnesses) are estimated and corrected for selection effects using a fiducial, and possibly misspecified, model[46]; then, they are correlated with galaxy properties obtained in a separate analysis (rarely considering the associated uncertainties[47]) or used in a cosmological fit.

A two-step approach cannot account for many important statistical effects. For instance, the ages of progenitors of type Ia supernova are not representative of the stellar populations of their hosts, and so regressing using the latter may be misleading. Moreover, standardization implicitly gives higher weight to more massive hosts, where a larger number of supernovae occur, an instance of Eddington bias[48,49]. Likewise, a preference for detecting (apparently) brighter objects biases the selected sample towards intrinsically brighter supernovae to a greater extent in dustier hosts, which are also more massive. Last, an apparent correlation—non-existent intrinsically—between supernova brightnesses and the properties of the hosts may be introduced by the use of photometric redshifts in the standardization process due to the mass–age–redshift degeneracy. Similarly, an apparent redshift dependence of supernova properties[50,51] might arise due to the evolution of their hosts, for example their dust content[52,53], metallicity or age[54].

In this paper, we present a framework for combined inference and galaxy-related standardization (CIGaRS) of type Ia supernovae and their hosts, which addresses all the above conceptual and methodological issues and can unequivocally disentangle intrinsic and extrinsic effects through physics-based forward modelling. We formulate our model in the context of simulation-based inference (SBI) (for overviews, see refs. 55,56), a modern suite of Bayesian inference techniques that leverage the flexibility of neural networks (NNs) to obtain posterior distributions, given only training examples from a forward simulator. SBI is seeing rapid adoption across disciplines and has previously been applied in studies of type Ia supernovae to analyse dust distributions from collections of light curves (LCs)[37,57] and to perform cosmological inference from future-sized datasets (with up to $10^5$ type Ia supernovae) in the presence of photometric-like redshift uncertainties[58] and arbitrary selection effects[49].

Our simulator (depicted graphically in Fig. 1 and fully elaborated in 'Unified forward modelling of galaxies and type Ia supernovae')

adopts the prescription for galaxy evolution from Prospector-β[59,60] and the Simple-BayeSN hierarchical model for LC summaries of type Ia supernovae[32]. It then applies dust extinction, cosmological distance, realistic measurement noise and sample selection to both. We model and infer the intrinsic correlations between (true) host metallicity, (true) progenitor age and (true intrinsic) supernova absolute magnitude, which are latent variables in the hierarchical model. In addition, we allow for a residual mass step and infer its size (in magnitudes) and location (host mass). Last, we include a self-consistent formulation of the occurrence of type Ia supernovae within galaxies through the delay-time distribution (DTD): the rate at which type Ia supernovae arise from the stellar population of their hosts based on their ages, which can be calculated from first principles for plausible progenitor scenarios for type Ia supernovae[61]. Our framework seamlessly integrates inference of the DTD under a suitable parameterization, thus offering a tool that can shed light on the much-debated formation mechanism of type Ia supernovae[62,63].

While insights into the DTD and the host–type Ia supernova connections are an important output of our analysis, we envisage its primary future application to be cosmological inference from the Legacy Survey of Space and Time (LSST), soon to commence at the Vera Rubin Observatory. Owing to the large spatial coverage and depth (18,000 deg$^2$ reaching redshift ~1) of its primary wide–fast–deep survey component, only ~1% of the ~10$^6$ supernova candidates expected over 10 years will have spectroscopic typing and redshift, and even with dedicated campaigns like 4MOST/TiDES[64], only up to 10% of the host galaxies will be followed up.

In view of this predominantly photometric scenario, we do not consider any spectroscopic observables or explicit redshift estimates, and we provide the inference network only with broadband photometry of the hosts and LC summaries for the type Ia supernovae (we do assume a perfect transient classification, noting that non-Ia contamination can be straightforwardly included in our framework). Owing to the physical modelling of galaxy evolution and the pooling of strength introduced by the joint analysis, our framework delivers state-of-the-art photometric redshift estimates as a by-product as well as considerably more stringent cosmological constraints than using supernova data alone (even with spectroscopic redshifts). We explore more thoroughly the implications of these findings for supernova cosmology with future datasets in a dedicated companion paper, focusing here on host–type Ia supernova connections and the DTD.

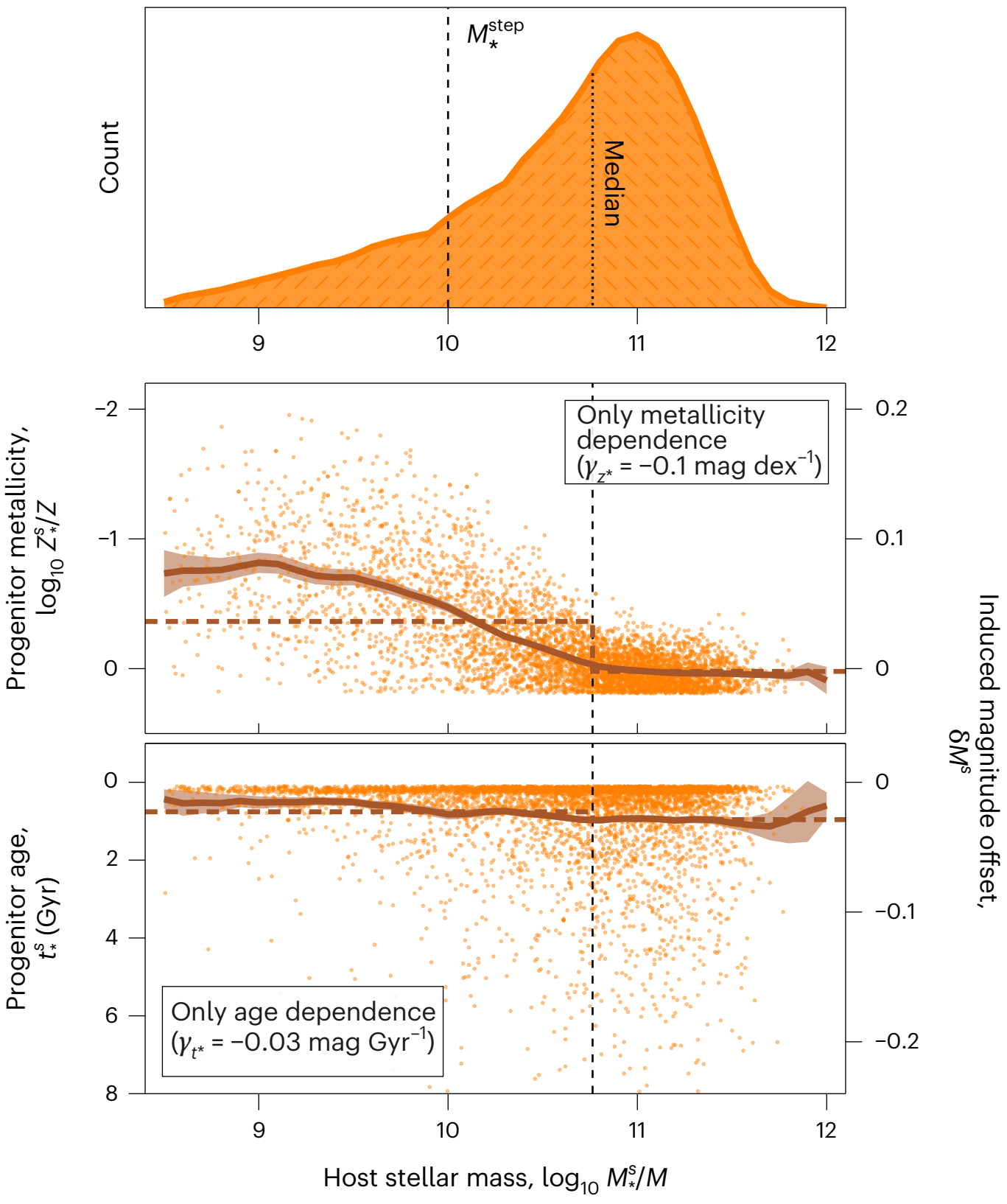

**Fig. 2 | Metallicity and age dependences as apparent 'mass steps'.** Top: distribution of the stellar masses of type Ia supernova hosts in our mock dataset, that is the galaxy stellar mass function weighted by the type Ia supernova rate within each galaxy (using DTD slope $b = -1.34$). Middle and bottom: true metallicities and ages of the progenitor stellar populations (left-hand $y$ axis) and the induced magnitude offsets of the type Ia supernovae $\delta M$ (right-hand $y$ axis) under the default linear correlation parameters ($\gamma_{Z_*}$ and $\gamma_{t_*}$), as indicated. Solid lines are a smooth moving average (with uncertainty due to sampling noise). Dashed lines depict the apparent step across the median host mass. The plots show the full population before detection and selection and across all redshifts. We also indicate the location ($M_*^{\rm step}$) of the extra step ($\Delta M = 0.05$) simulated in our mock data.

## Results

To examine the phenomenology and demonstrate inference with our unified model, we generate a mock dataset $D_0$, with global parameters as listed in Extended Data Table 1 and object-specific (latent) quantities sampled from their hierarchical priors. We scale the counts to be representative of the current flagship Dark Energy Survey (DES), which contains 1,635 photometrically classified supernovae[65]. In our particular $D_0$, the number of selected objects $N_{\rm sel,0} = 1{,}578$.

### Predictions from the unified forward model

**Physical host–type Ia supernova correlations and induced stellar-mass dependence.** Our forward model includes an explicit dependence of the magnitude offsets of type Ia supernovae ($\delta M^s$) on the true progenitor metallicity ($Z_*^s$) and age ($t_*^s$): see equation (10). We plot those (separately) in Fig. 2 (with correlation parameters $\gamma_{Z_*}$ and $\gamma_{t_*}$ chosen consistently with observations[8,12] and indicated in the respective panels) against the stellar mass of the hosts. Owing to well-known trends in galaxy evolution (more massive galaxies are, in general, older and more metal-rich), the brightness of type Ia supernovae appears to depend on stellar mass. These are incorporated in Prospector-β through the use—with appropriate scatter—of an empirical star-formation history (SFH)[66] and mass–metallicity relation[67], and they give rise to a crucial difference: whereas a metallicity correlation ($\gamma_{Z_*} \neq 0$) induces a (smoothed) step across $M_* \approx 10^{10}$, a progenitor-age correlation ($\gamma_{t_*} \neq 0$) manifests as an approximately linear trend with $\log M_*$. Moreover, the distributions of magnitude offsets at a given stellar mass, $p(\delta M \mid M_*)$, also differ: although the metallicity shows a Gaussian spread, progenitor ages are highly skewed towards younger populations due to the fast-decaying DTD (we adopt a power law with slope $b = -1.34$; ref. 68).

Owing to these differences, in principle, the two scenarios can be distinguished using stellar-mass estimates alone, which are easy to obtain: see our results in section 'Object-specific parameters'. However, as we show next, this task is complicated by the presence of colour- and stretch-related variations and residual scatter (~0.3 mag in total), as well as uncertainties both in the dependent ($\delta M^s$) and independent variables ($M_*^s$, $Z_*^s$ and $t_*^s$). Finally, note that splitting at the median mass—as customary—results in a diminished step (dashed lines) as both subsets are 'pulled' towards the peak of the $M_*$ distribution (~$10^{11}\,M_\odot$). This is an instance of Eddington bias[48,49].

**Redshift distribution and Hubble diagram.** A key outcome of our unified approach is the predicted redshift distribution of type Ia supernovae, which results from the combination of galaxy evolution (according to Prospector-β) and a DTD-based occurrence model.

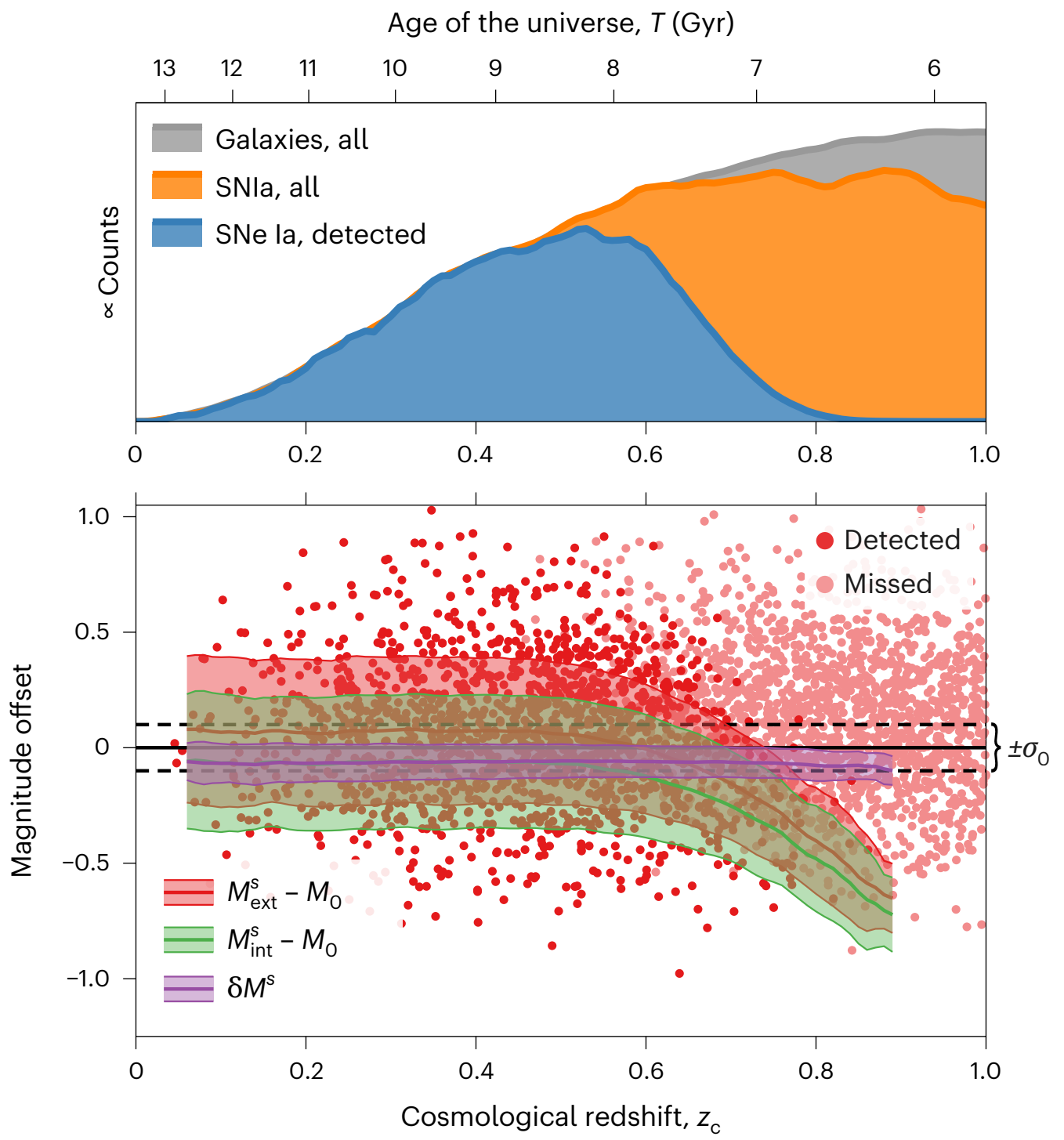

**Fig. 3 | Hubble diagram.** Top: redshift distributions of galaxies (according to Prospector-β), type Ia supernovae within them (as predicted by our DTD-based model) and detected type Ia supernovae (using our LSST-inspired selection function). While the latter two are on the same scale, the galaxy counts have been renormalized for clarity. Bottom: offsets of the extrinsic (dust-affected) and intrinsic absolute magnitudes ($M^s_{ext}$ and $M^s_{int}$) from the standard $M_0$ and the host- and progenitor-related $\delta M^s$ as functions of redshift. Points represent $M^s_{ext}$ from our mock data (undetected supernovae are shown in a lighter shade). The shaded areas indicate the mean ± standard deviation—for detected supernovae—of the respective quantities binned by $z_c$. The dashed lines delineate the irreducible and residual scatter with standard deviation $\sigma_0$ intrinsic in the type Ia supernova population.

At low cosmological redshifts (up to $z_c \approx 0.3$), it is qualitatively consistent with a power-law volumetric rate with exponent 1 to 1.5, as inferred by Dilday et al.[69] from Sloan Digital Sky Survey data and widely used in the literature (for example, in PLAsTiCC[70]). Above $z_c \approx 0.4$, the rate starts decreasing due to the decline in the galaxy population. Finally, at $z_c \gtrsim 0.6$, corresponding to a Universe of age ≲8 Gyr, the frequency of type Ia supernova occurrence within hosts decreases, as potential hosts are smaller (as there is less time for star formation) and younger (thus, reducing the range of the DTD). This effect is compounded with the drop in detection efficiency, and the result is that—for our selection function, which mimics the LSST wide–fast–deep survey ('Detection and selection of type Ia supernovae')—no type Ia supernovae are detected above $z_c \approx 0.8$.

In the bottom panel of Fig. 3, we show the different absolute magnitude offsets of the type Ia supernovae included in our forward model. These would be the residuals from a perfectly specified cosmological model and standard magnitude $M_0$ in a traditional analysis. The dominant source of scatter is standardizable correlations with stretch and colour (~0.26 mag). At high redshift, the sample exhibits a typical Malmquist bias[71,72] towards brighter supernovae (downwards trend) with a reduction of observed scatter due to the removal of dim objects. The effect of dust is also systematic. It dims (shifts upwards) the extrinsic magnitudes $M^s_{ext}$ (red-shaded band) with respect to $M^s_{int}$ (green-shaded band) by ~0.13 mag. Finally, we show the intrinsic host-related magnitude offset ($\delta M^s$, purple-shaded band), which includes metallicity and age correlations, as visualized in Fig. 2, and a residual mass step of 0.05 mag across $10^{10}\,M_\odot$. These effects lead to an offset due to the dominance of high-mass (and hence, higher-metallicity and older) galaxies, but their sizes are small in comparison with the above (on the order of the irreducible random scatter: $\sigma_0 = 0.1$ mag), making the host–type Ia supernova connections difficult to infer.

Traditional analyses need to incorporate all these effects in an intricate likelihood that attempts to disentangle intrinsic and extrinsic influences, scatter from measurement uncertainty, trends from selection biases to derive accurate constraints on the cosmological parameters, host–type Ia supernova connections, and the DTD. In our SBI framework, on the other hand, it is sufficient to include them in the forward model, and the NN learns from training data how to solve the full inverse problem.

## Simultaneous SBI

In this section, we demonstrate the ability of CIGaRS to obtain—simultaneously—marginal posteriors for all parameters in the hierarchical model, given only LC summaries of type Ia supernovae (stretch, apparent brightness, colour at peak and the associated fitting uncertainties) and LSST-like ugrizy photometry of the hosts. We use truncated marginal neural ratio estimation (TMNRE)[73,74] to train a conditioned deep set++ network[49,75,76] (Fig. 4) and apply it to the simulated dataset presented above. With minimal fine-tuning, we then scale to a tenfold larger mock catalogue of ~16,000 type Ia supernovae and their hosts (expected to be discovered in a month's worth of wide–fast–deep survey observations). Details of the network architecture and the iterative training procedure are presented in section 'Set-based TMNRE for hierarchical models'.

**Global and population parameters.** Figure 5 shows the final posteriors for all global and population parameters. All groups are significantly constrained with respect to their initial priors, and the results are fully consistent with the values used to simulate the mock data.

We constrain the power-law slope of the DTD, $b$, to within ±0.05, comparable with the uncertainty in the DES analysis[77] of 809 objects, which used spectroscopic redshifts. We also recover the parameters of all three kinds of intrinsic host–type Ia supernova dependences that we consider: correlations with the progenitor metallicity or age and the size and location of an additional mass step. Importantly, we can disentangle a metallicity from an age dependence even when both are present (as evident from the uncorrelated posterior in the $\gamma_{Z_*}$–$\gamma_{t_*}$ plane). By contrast, a metallicity dependence is harder to distinguish from a mass step (hence the strong a posteriori correlation in the $\Delta M$–$\gamma_{Z_*}$ plane), as could be anticipated from our exploration of the model predictions (Fig. 2).

Finally, we present marginal posteriors for the cosmological parameters (dark matter and dark energy densities, $\Omega_{m0}$ and $\Omega_{\Lambda0}$) and compare them with an idealized analysis in which all the latent parameters (redshifts, stretches, intrinsic colours, dust and host-related magnitude offsets) are exactly known, and the only sources of uncertainty are the irreducible magnitude scatter of $\sigma_0 = 0.1$ mag and measurement noise. Even this unrealistic scenario, which would at minimum require expensive spectroscopic observations, delivers cosmological constraints (in terms of area in the $\Omega_{m0}$–$\Omega_{\Lambda0}$ plane) that are only 3 times stronger.

However, the main advantage of our fully general and realistic approach is that it can be applied to much larger photometric-only datasets. To illustrate the improvement we expect from including all discovered type Ia supernovae rather than only those with spectroscopic follow-up (about 10% at most with LSST and 4MOST/TiDES), we also show the posteriors obtained from our ×10 larger mock dataset. We verify that the precision of our principled and self-consistent analysis scales as expected with the survey size. The one-dimensional credible intervals are $\sqrt{10} \approx 3$ times narrower, and the areas of two-dimensional regions shrink 10 times. Thus, CIGaRS will allow robust conclusions to be drawn about the progenitors of type Ia supernovae and their environmental dependencies, which has never been achieved before, while also improving cosmological constraints ~4 times over current methodologies that rely on spectroscopy.

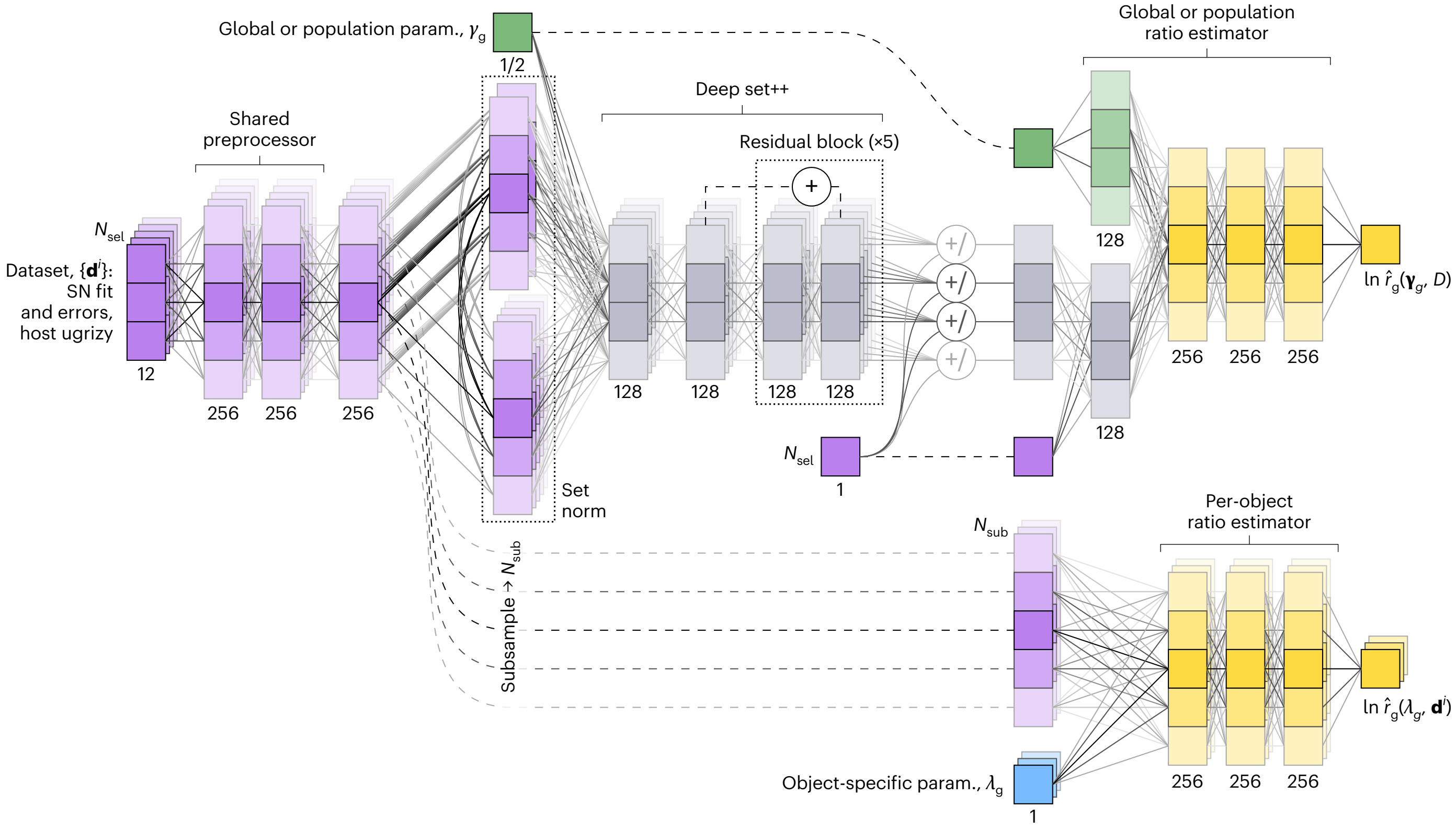

**Fig. 4 | Our inference network.** The network combines a conditioned deep set++ (using residual connections and set normalization)[49,75,76] with standard feed-forward ratio estimators to infer simultaneously the global and population parameters in our forward model and all object-specific properties from a catalogue containing summary statistics for type Ia supernovae (with uncertainties) and photometry of their hosts.

**Object-specific parameters.** Marginal posteriors for the parameters of all 1,578 hosts and type Ia supernovae—derived simultaneously with the global results—are presented in Fig. 6 and Extended Data Fig. 1. The true values are recovered well across the full ranges, and although our results for the stretch and colour parameters are largely driven by the (noisy) direct 'measurements' ($\hat{x}^s \pm \hat{\sigma}^s_x$ and $\hat{c}^s \pm \hat{\sigma}^s_c$) included in the data, they are also informed by the respective population distributions, which reduces the uncertainties when the observational errors are large. This is an instance of Bayesian shrinkage. For the host stellar mass and metallicity, our uncertainties match those from previous applications of Prospector[59,60]. When the 'signal' is intrinsically low, for example for low metallicities, our results revert to the prior, which ensures proper Bayesian uncertainty propagation. Moreover, photometric redshifts are inferred with exceptionally high precision: a median posterior standard deviation ~0.01, no outliers with $|\Delta z| > 0.1$ and no signs of a systematic bias, owing to (1) the Prospector-β priors informed by galaxy evolution, which break the age–mass–redshift degeneracy[60] and (2) the extraction of redshift information from supernova (and host) brightnesses, given the globally inferred cosmological model. This is on par with analyses of ultraviolet–optical–infrared data with much wider coverage[78] and represents a considerable improvement over the expected scatter of ~0.04 from LSST photometry alone[79]. Despite using only summarized supernova observables, our results already surpass the state of the art with full LC and host photometry. See Chen et al.[80], who achieved a photo-$z$ scatter of ~0.02 for DES.

## Conclusion and outlook

We have presented CIGaRS, a unified and fully self-consistent framework for the rigorous Bayesian analysis of summary statistics of type Ia supernovae and the photometry of their hosts. It is based on a forward simulator that combines two hierarchical models, Prospector-β and Simple-BayeSN, via a prescription for the occurrence of type Ia supernovae (according to a DTD convolved with non-trivial SFHs) and physical connections between their brightnesses and progenitor metallicity or age. CIGaRS qualitatively reproduces the observed type Ia supernova rate with redshift and the distribution of host stellar masses and makes detailed predictions about the apparent stellar-mass dependence induced by the underlying physical relations. A metallicity correlation resembles a smoothed step function across $10^{10}\ M_\odot$ with symmetric Gaussian scatter, whereas an age correlation leads to a linear trend with stellar mass and highly skewed residuals.

We have also demonstrated simultaneous neural SBI of the DTD parameters, brightness versus metallicity and age correlations, a residual mass step, and cosmology purely from LSST-like host photometry and summaries of type Ia supernovae (brightness, stretch and colour) of up to ~16,000 objects. Our approach implicitly handles redshift estimation, host-related standardization and selection effects while propagating all uncertainties, and thus, it remains accurate, precise and scalable. Moreover, due to the unified hierarchical modelling and combination of supernova and host data, we achieve robust photometric redshift estimates with state-of-the-art scatter and cosmological results that rival spectroscopic supernova-only analyses. In a companion work, we investigate the ability of CIGaRS to also standardize the host photometry and the effect that this has on the cosmological constraints.

We note, finally, that the present analysis is performed within a constrained—albeit fully self-consistent—modelling framework, which may not exactly represent reality. For example, we have fixed (based on previous studies) foundational details of the Prospector model, like the stellar initial mass function, stellar isochrones and the intergalactic dust-extinction law. The systematic effect of such choices can be easily examined a posteriori with SBI by quickly analysing various simulated datasets. We show in Extended Data Figs. 2 and 3 one example of replacing the KC13 law[81] with F99[82]. The resulting bias is imperceptible at the individual-object level but significantly affects the pooled global-parameter inference, especially as the data grow larger and more constraining. Nevertheless, with SBI, we can properly account for such 'known unknowns' by treating them as nuisance variables, like the (non-)presence of the various host–type Ia supernova connections we already consider. Concretely, by training the network on mock data with a variety of initial mass functions, isochrones, dust laws and so, it can infer the correct models if they are clearly discernible from the data and automatically include the appropriate 'systematic' uncertainty associated with the modelling choice.

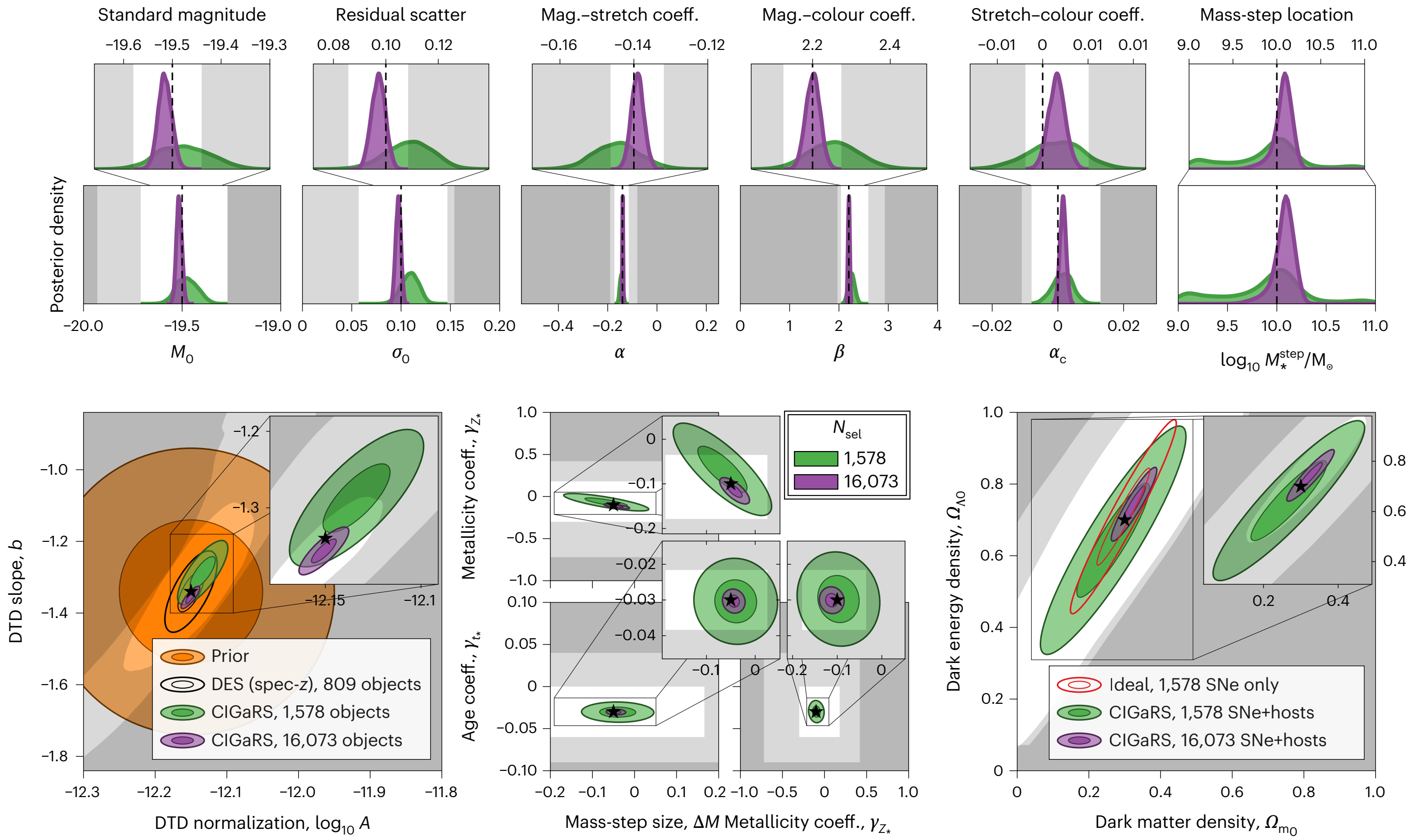

**Fig. 5 | Inference results from the two mock datasets.** Top: marginal posterior densities. Bottom: 1$\sigma$ and 2$\sigma$ contours (with credibility 39% and 86%, respectively). True values are indicated by dashed lines and stars. For all parameters, the prior is uniform over the plotted range, except for the DTD, where it is an uncorrelated normal distribution (orange). Grey-shaded areas indicate successive stages of prior truncation. For comparison, we show the DTD constraints (shifted to our true value) from the DES[77], using 809 objects with spectroscopic redshifts, and an idealized cosmological analysis (given the ground-truth values for all object-specific latent parameters, which are not available in reality) of only the observables of type Ia supernovae from our mock data (excluding hosts). coeff., coefficient; Mag., magnitude; SNe, supernovae; spec-*z*, spectroscopic redshift.

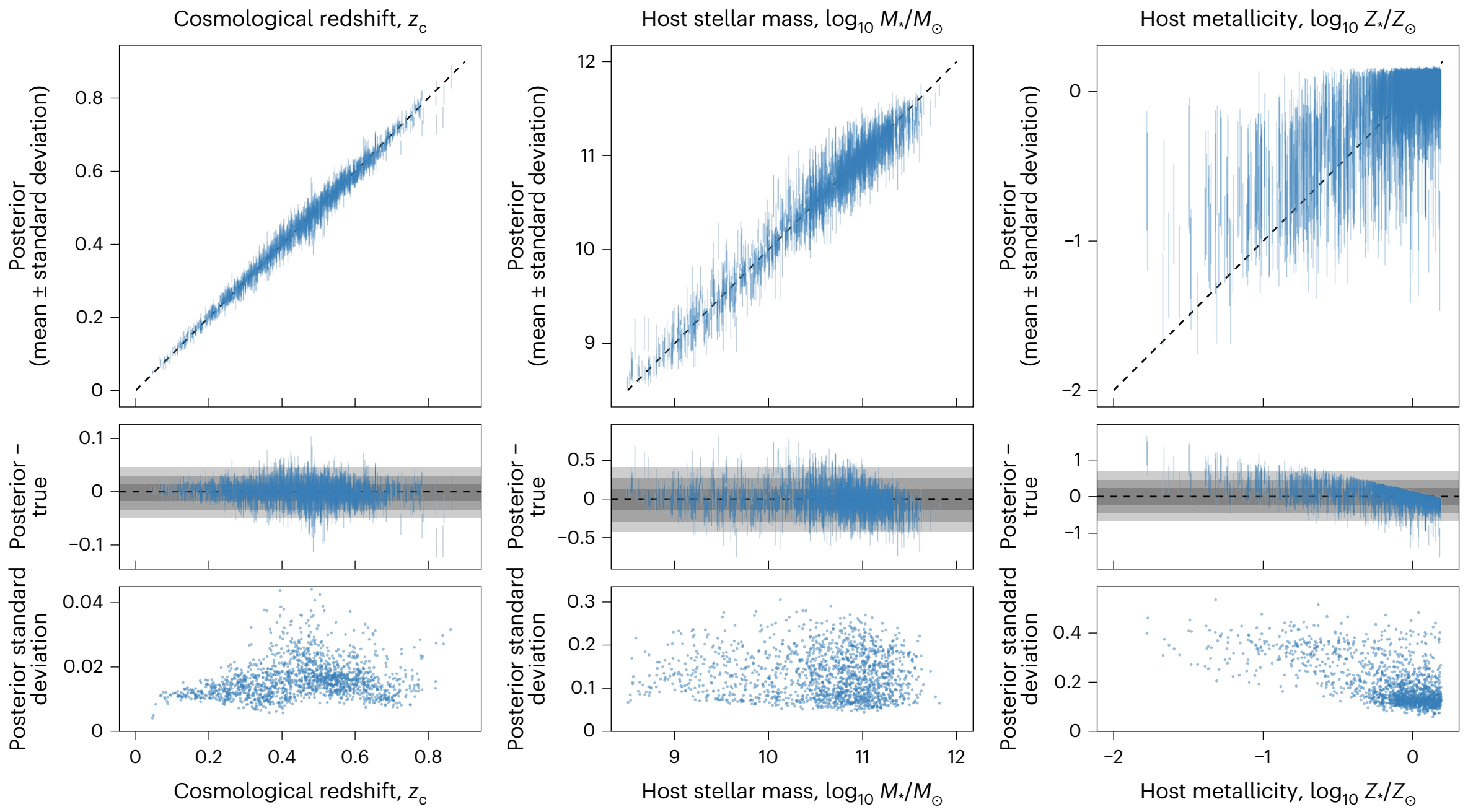

**Fig. 6 | Inference results for host-related parameters.** Inference results for the 1,578 objects in our mock data, obtained simultaneously with the global constraints. Everywhere, the horizontal axis is the true parameter value in the simulation. Top: all 1,578 marginal posteriors (mean ± standard deviation). Middle: residuals from true values. The shaded bars represent the distribution (1, 2 and 3 standard deviations) of the posterior means. Bottom: the posterior uncertainties (standard deviations). We show only $z_c$, $M_*$ and $Z_*$ as they are relevant to the current study, but we infer the rest of the object-specific parameters as well (Extended Data Fig. 1), as this helps train the shared preprocessor. See section 'Set-based TMNRE for hierarchical models'.

Further extensions of the model that we envisage include the spatial localization of the stellar populations (and, hence, of the supernovae) and of interstellar dust, the peculiar velocities of the hosts and of the supernovae within them (which will allow inference of cosmological parameters controlling large-scale structure), the presence of non-Ia transients, their detection and characterization directly from a time series of telescope images, and the incorporation of spectroscopic data, where available. Notably, our inference framework can be applied to an arbitrarily complicated simulator and collections of observables without major modifications, while retaining its scalability, and we expect it to be instrumental in maximizing the scientific insight extracted from upcoming vast and detailed LSST observations.

# Methods

## Unified forward modelling of galaxies and type Ia supernovae

CIGaRS is a unified forward model for type Ia supernova cosmology. It is a diptych (Fig. 1) of two state-of-the-art Bayesian hierarchical simulators for photometric galaxy observations ($\mathbf{d}^h$) and summary statistics of type Ia supernovae ($\mathbf{d}^s$), joined on three levels:

1. by the probability of occurrence (equivalently, number $N_{\mathrm{SN}}^h$) of type Ia supernovae in a given host $h$, based on the intrinsic properties of the latter, $\mathbf{g}^h$;
2. by the distribution of the intrinsic properties $\boldsymbol{\lambda}^s$ of the supernovae, which also depend on the $\mathbf{g}^{h(s)}$ of their respective hosts;
3. by extrinsic effects of the host environment (for example, dust) and other properties (for example, cosmological redshift and peculiar velocity) on supernova light as it travels towards the observer.

The former two, which affect $\boldsymbol{\lambda}$, can be considered causal connections, whereas the latter, which affects only the observables and data $\mathbf{d}$, are coincidental. One last connection can arise during data reduction (for example, subtraction of the host light). We plan to implement this in a future extension that also implements the image analysis procedure of transient surveys. In the following, we describe in detail all components of CIGaRS (see also the full hierarchy in Fig. 1 and the list of parameters in Extended Data Table 1) and discuss the planned improvements, which are mainly related to the intrahost localization of the supernovae.

**Host model.** A comprehensive treatment of the photo-spectral evolution of galaxies—see the pioneering studies using the MOPED method[83–85] and, for example, Mo et al.[86] for a review—is beyond the scope of this work. At a high level, the intrinsic properties of a galaxy $h$, such as its SFH, metallicity ($Z_*^h$) and interstellar dust content, are largely determined by the total stellar mass of the galaxy ($M_*^h$) and its cosmological redshift ($z_c^h$), the latter being a proxy (given a cosmological model) for the age of the Universe ($T^h$) at the time the galaxy's light was emitted. To represent the intrinsic correlations inherent in galaxy formation, we adopt an off-the-shelf Bayesian framework for galaxy photometry: Prospector[87], as updated in Prospector-β[60].

**SFH and chemical enrichment: Prospector-β.** Prospector encodes a non-trivial SFH by discretizing it into seven time bins (the optimal number for describing moderate signal-to-noise observations[88]) spanning $[0; T^h]$: the first two are fixed to [0 Myr; 30 Myr] and [30 Myr; 100 Myr], the next four logarithmically spaced within $[100\ \mathrm{Myr}; 0.85T^h]$, and the last covering $[0.85T^h; T^h]$. The total mass of a galaxy is then distributed into seven stellar populations that represent the stars born during each history bin:

$$\mathbf{SFH}^h \equiv \left[\mathrm{SFH}^{h,j}\right]_{j=1}^{7}, \quad \text{with} \quad \sum_{i=1}^{7} \mathrm{SFH}^{h,j} = M_*^h. \tag{1}$$

In each forward realization, $\mathbf{SFH}^h$ is sampled from the prior $p(\mathbf{SFH}^h | M_*^h, T^h)$ introduced in Prospector-β (Section 3.3 of ref. 60). The model adopts a non-parametric continuity prescription[89] whereby the ratio of SFRs in adjacent bins exhibits a random scatter around the value dictated by the average ('cosmic') SFR density[66], taking into account also the empirical observation that high-mass galaxies form earlier.

For a given $\mathbf{SFH}^h$, and assuming that all stars in a given population were born at approximately the same time ($t_*^{h,j}$, taken at the (linear) centre of the respective bin), the process of stellar population synthesis (SPS) produces the integrated light (before extinction) from all stars in a galaxy by convolving the emission spectra of single stellar populations with the SFH.

SPS also requires a prescription for the galactic chemical enrichment history: the metallicities of all its stellar populations. Prospector assumes a single metallicity within a galaxy ($Z_*^h$), determining it (with appropriate scatter) from the total stellar mass, as in Gallazzi et al.[67]. This has also been shown to be appropriate for modelling the progenitors of type Ia supernovae[90], which is our ultimate goal.

**Host dust extinction.** The light emitted by stars is then extinguished (and reddened) by dust within the host. Prospector considers separate birth-cloud and diffuse dust components. The former affects only the youngest stellar population (acting on the emission only from the first time bin) and has a fixed wavelength dependence proportional to $\lambda^{-1}$ and optical depth $\tau_1^h$. Extinction due to diffuse (interstellar) dust is described by the KC13 law (ref. 81, following refs. 91,92) with shape parameter $\delta^h$ and optical depth $\tau_2^h$ related to the column density of dust along the line of sight (as it represents the total extinction in the (rest-frame) V-band, we will label it $A_V^h$ instead).

In Prospector-β, the dust parameters are a priori independent from other galaxy properties; however, many studies have found a posteriori relations from large galaxy surveys. We enforce the empirical relations of Alsing et al. (equations (14) and (15) in ref. 93):

$$\tau_2^h \equiv A_V^h \sim \mathcal{N}\left(0.2 + 0.5\,\mathrm{ReLU}\left(\log_{10}\left[\mathrm{SFR}^h\ (M_\odot\ \mathrm{yr}^{-1})\right]\right), 0.2^2\right), \tag{2}$$

$$\delta^h \sim \mathcal{N}\left(-0.095 + 0.111A_V^h - 0.0066(A_V^h)^2, 0.4^2\right), \tag{3}$$

where the SFR is derived from the most recent (current) component of $\mathbf{SFH}^h$: $\mathrm{SFR}^h \equiv \mathrm{SFH}^{h,1}/30\ \mathrm{Myr}$. Prospector then sets the optical depth of the birth-cloud dust according to $\tau_1^h/\tau_2^h \approx \mathcal{N}(1, 0.3^2)$.

Importantly, the above description of dust extinction is global, that is averaged over the spatial light distribution of the galaxy, as the host photometry is usually integrated. It may, thus, not faithfully represent the local extinction law or the optical density of dust at any particular location within the galaxy, for example in the proximity of a given type Ia supernova. The spatial distribution of dust properties is a topic of active research (for example, refs. 94–97), with mounting evidence for its effect on the standardization of type Ia supernovae[98,99]. We will address this in a future refinement of the model, in coordination with the dust effects applied to supernova light.

**Photometry and detection of hosts.** Given the above quantities (and parameters that describe the presence and emission of interstellar gas and an active galactic nucleus, which we randomly sample from the Prospector-β priors as they are not directly relevant to studies of type Ia supernovae), we compute the rest-frame spectral energy distribution—more accurately, the spectral flux—$\Phi^h(\lambda_r)$ using the NN-based SPS emulator Speculator-α[100]. The speed-up this brings over traditional (from first principles) SPS is crucial for simulating the large amount of training data we need.

We then apply a redshift to $\Phi^h$ (in this study, we assume no peculiar velocities, $z_c{}^h = z^h = z^s$, as the fraction of objects in our simulations and mock data with $z_c \lesssim 0.02$, where the distinction would be important, is negligible) and apply a cosmological (transverse co-moving) distance $D^h \equiv D(z_c^h, \mathcal{C})$, where $\mathcal{C}$ are the cosmological parameters, to convert it to the observer-frame spectral flux density:

$$F^h(\lambda_o) \equiv \frac{\Phi^h(\lambda_o/(1+z^h))}{(1+z^h)^3 4\pi (D^h)^2}, \tag{4}$$

where $\lambda_o$ is the observer-frame wavelength. In principle, at this stage we need to account for extinction in intergalactic space[101,102] and in the Milky Way. The former is a small effect (~0.015 mag at $z_c = 0.6$) and is rarely addressed explicitly in supernova cosmology (see, for example, Section 4.4.7 in ref. 29), whereas the latter can be accounted for by using detailed multi-wavelength attenuation maps of the Milky Way[103–106]. Hence, in this work, we ignore the two effects and assume that the observations have been appropriately corrected.

Thus, we directly integrate (numerically) $F^h$ within the Rubin ugrizy passbands, convert to AB magnitudes and apply 0.01 mag Gaussian noise (a conservative estimate for the Rubin Observatory's absolute photometric calibration[107]) to arrive at the simulated galaxy data $\mathbf{d}^h$, a vector of length 6. Finally, to simulate detection, that is the indicator variable $S^h$, we apply a simple magnitude cut by stipulating that all six measured magnitudes must be brighter than the respective $5\sigma$ detection thresholds expected for LSST 10-yr co-added imaging[108].

The description above treats galaxies as isotropically emitting point sources, whereas in reality, measuring their brightnesses requires accounting for inclination and de-blending of overlapping objects, which—to first order—would inflate the error and uncertainty beyond the photometric calibration floor we have assumed. To account for higher-order effects (for example, complicated biases due to inclination, blending or selection effects arising from the complex procedure for detecting and selecting objects for inclusion in a galaxy catalogue), one would need to process the raw image data. Owing to the flexibility of neural SBI, this can be achieved through high-fidelity simulations (see, for example, refs. 109–113) to which the same data-reduction procedure has been applied as to the real data.

**Occurrence of type Ia supernovae.** The first connection between galaxies and type Ia supernovae (or any other transient) is the emergence of the latter from the stellar populations of the former, which we describe through the DTD: the rate (per unit rest-frame time) of occurrence of type Ia supernovae from a given stellar population (per unit stellar mass within it), as a function of the age of the population $t_*$. We adopt a power-law parameterization:

$$\mathrm{DTD}\,(t_*) = A \times (t_*/\mathrm{Gyr})^b \times \mathrm{M}_\odot{}^{-1}\,\mathrm{yr}^{-1}, \tag{5}$$

which is guided by the theory of binary systems that decay through the emission of gravitational waves (for which $b = -1$), with independent priors on the normalization and slope from observational constraints[68]:

$$\log_{10} A \sim \mathcal{N}\left(-12.15, 0.1^2\right), \tag{6}$$

$$b \sim \mathcal{N}\left(-1.34, 0.2^2\right). \tag{7}$$

To avoid an infinite total rate, we set the DTD to zero for $t_* < 0.1$ Gyr (that is, for the two youngest stellar populations in Prospector), corresponding to the minimum time for the creation of a carbon-oxygen white dwarf[114,115]. This physical prescription resolves the strong a posteriori correlation between the cutoff time and $A$, which has previously prevented direct inference of the former[77]. In future analyses, we plan to relax this assumption and include a mixture of formation channels (for example, ref. 116), modelling their DTDs from first principles.

Given the DTD, we calculate the number of type Ia supernovae that arise (on expectation) from the stellar population $j$ of galaxy $h$ (recall that we assumed all stars within it have the same age $t_*^{h,j}$) during a survey of (observer-frame) duration $\mathcal{T}$:

$$\left\langle N_{\mathrm{SN}}^{h,j}\right\rangle = \frac{\mathcal{T}}{1+z^h} \times \mathrm{SFH}^{h,j} \times \mathrm{DTD}\,(t_*^{h,j}). \tag{8}$$

We then iterate (in parallel on a GPU) over all stellar populations of all galaxies, that is over all $h$ and $j$, to generate

$$N_{\mathrm{SN}}^{h,j} \sim \mathrm{Poisson}\left(\left\langle N_{\mathrm{SN}}^{h,j}\right\rangle\right) \tag{9}$$

supernovae. We sample their ages, $\{t_*^{h,j,k}\}_{k=1}^{N_{\mathrm{SN}}^{h,j}}$, uniformly within the time bin associated with population $h$ and $j$. To simplify notation, we uniquely label each supernova with $s \in \{1, \ldots, \sum_{h,j} N_{\mathrm{SN}}^{h,j}\}$ and keep track of the host $h(s)$ of each, so that we can associate the relevant intrinsic galaxy properties: $z_c^s \equiv z_c^{h(s)}$, $Z_*^s \equiv Z_*^{h(s)}$ and $M_*^s \equiv M_*^{h(s)}$ and host observations $\mathbf{d}^{h(s)}$. We note that this assumes a perfect host association, which is often not the case in real observations, and we ignore the extra information from instances of several supernovae arising in the same host (see, for example, ref. 117); we plan to make improvements in these respects by extending the simulator towards detection and characterization of the transients directly from telescope images.

Finally, it is also possible to keep track—in the simulator—of the stellar population $j(s)$ from which each supernova has arisen and associate the corresponding properties, if these differ from one stellar population to another within the same host (for example, to represent a fully realistic enrichment history through different $Z_*^{h,j}$). Similarly, we can simulate the spatial distribution of stellar populations that give rise to metallicity and age gradients (for example, refs. 118,119) and differences in dust extinction (for example, refs. 94–97). We defer an expansion of the simulator within the hosts to future work.

**Model for type Ia supernovae.** Once we have determined how many type Ia supernovae have occurred during a survey, we proceed to simulate their observables. This follows a similar path, going from the intrinsic properties derived from a population model informed by the host of each object, through extinction, redshift and distance to noisy measurements and detection or sample selection.

**Causal host–type Ia supernova connections.** The cornerstone of CIGaRS is an explicit parameterized dependence of the intrinsic properties of type Ia supernovae on the characteristics of their host or of their progenitor stellar population: in general, $p(\boldsymbol{\Lambda}^s|\mathbf{g}^{h(s),j(s)}, \boldsymbol{\gamma})$, where $\boldsymbol{\gamma}$ are global parameters. Our choice of which links to include is motivated by observational evidence[4,6–12] and takes the form of a host-dependent magnitude offset $\delta M^s$, but we could similarly introduce, for example, a parameterized host-dependent distribution of $x_{\mathrm{int}}^s$ (ref. 26).

As the physical source of $\delta M^s$ is not yet established, we allow for several possibilities (in isolation or combination)—namely, metallicity $Z_*^s$ and progenitor age $t_*^s$—with correlation coefficients, respectively $\gamma_{Z_*}$ and $\gamma_{t_*}$, whose priors include 0 and wide ranges of positive and negative values. In keeping with current practice and as a 'catch-all' parameter, we also allow for an additional (or residual) 'mass step' $\Delta M$ with a similar wide prior and a free (inferred) stellar-mass location $M_*^{\rm step}$. The full intrinsic host connection is, therefore:

$$\delta M^s = \underbrace{\gamma_{Z_*} \times \log_{10} Z_*^s/Z_\odot}_{\text{metallicity}} + \underbrace{\gamma_{t_*} \times t_*^s}_{\text{age}} + \underbrace{\Delta M \times \mathbb{I}(M_*^s > M_*^{\rm step})}_{\text{ad hoc mass step}}, \quad (10)$$

where $\mathbb{I}$ returns 1 or 0 if its argument is true or false, respectively, and $\gamma_{Z_*}, \gamma_{t_*}, \Delta M, M_*^{\rm step} \in \boldsymbol{\gamma}$ are global parameters. Inference of the presence and strength of the respective effects is performed by examining their posterior(s) or performing a Bayesian model selection, which is most conveniently and rigorously achieved through simulation-based methods[37]. Moreover, with SBI, all underlying correlations are accounted for and marginalized over when inferring the cosmology, which ensures the robustness of dark energy inference against systematics arising from the host–type Ia supernova dependences.

**Intrinsic properties and dust extinction of type Ia supernovae: Simple-BayeSN.** Bayesian hierarchical modelling replaces empirical standardization with a priori correlated latent parameters sampled from hierarchical priors[32,120]:

$$\text{stretch: } x_{\rm int}^s \sim \mathcal{N}(\mu_x, \sigma_x^2), \quad (11)$$

$$\text{colour: } c_{\rm int}^s \sim \mathcal{N}\left(\mu_c + \alpha_c x_{\rm int}^s, \sigma_c^2\right), \quad (12)$$

$$\text{absolute B-band magnitude: } M_{\rm int}^s \sim \mathcal{N}\left(M_0 + \alpha x_{\rm int}^s + \beta c_{\rm int}^s + \delta M^s, \sigma_0^2\right) \quad (13)$$

where $\mu_x, \sigma_x, \mu_c, \sigma_c, M_0, \sigma_0, \alpha, \alpha_c, \beta \in \boldsymbol{\gamma}$ are population parameters assigned fixed hyper-priors, as listed in Extended Data Table 1.

Like starlight, supernovae are affected by the dust surrounding them and along the line of sight, that is, in intergalactic space and in the Milky Way. We will ignore the latter two, assuming that they have been perfectly corrected for, and adopt the Bayesian formulation of host dust extinction from Simple-BayeSN[32], which acts on the intrinsic parameters introduced in equations (11) to (13) to obtain their 'extrinsic' versions:

$$x_{\rm ext}^s = x_{\rm int}^s, \quad (14)$$

$$c_{\rm ext}^s = c_{\rm int}^s + E_{B-V}^s = c_{\rm int}^s + A_V^s/R_V^s, \quad (15)$$

$$M_{\rm ext}^s = M_{\rm int}^s + A_B^s = M_{\rm int}^s + A_V^s(R_V^s + 1)/R_V^s, \quad (16)$$

where $E_{B-V} \equiv A_B - A_V$ is the selective extinction, with $A_B$ the total extinction in the rest-frame B-band (in which the peak magnitude of type Ia supernovae is typically standardized) and $R_V \equiv A_V/E_{B-V}$. Note that the colour–magnitude standardization coefficient $\beta$ in equation (13) refers only to $c_{\rm int}$ rather than the dust-affected $c_{\rm ext}$.

Ideally, in a unified model, one would coordinate the extinction applied to type Ia supernovae with the dust properties of the host. However, due to the complicated distribution of dust within galaxies (for example, refs. 94–97) and the non-trivial attenuation effects of dust, which include not only extinction but also scattering of galactic and supernova light into the line of sight[31], the simple approach of adopting the host-global $A_V^h$ and $R(\delta^h)$ predicts an order of magnitude larger optical depths than empirically observed in supernova data. We will explore the effect of dust localization in a further study, and here we adopt instead the host-independent dust populations from Simple-BayeSN:

$$R_V^s \sim \mathcal{N}(\mu_R, \sigma_R^2), \quad (17)$$

$$A_V^s \sim \text{Exponential}(1/\tau), \quad (18)$$

with $\mu_R, \sigma_R, \tau \in \boldsymbol{\gamma}$ global parameters (Extended Data Table 1).

**Cosmological distance.** The extrinsic absolute (rest-frame B-band) magnitude $M_{\rm ext}^s$ is then transformed into an apparent (still rest-frame B-band) magnitude $m^s$ through the usual distance-modulus relation:

$$m^s = M_{\rm ext}^s + \mu^s \quad \text{with} \quad \mu^s = \mu(z_c^s, \mathcal{C}), \quad (19)$$

where $z_c^s$ is the cosmological redshift of the supernova, and $\mathcal{C}$ are the cosmological parameters. In this study, we use $\Lambda$-cold dark matter, which is described by the present-day dimensionless densities of cold dark matter and dark energy in the form of a cosmological constant: $\mathcal{C} \equiv [\Omega_{\rm m0}, \Omega_{\Lambda 0}]$. Just as when modelling galaxies, we disregard peculiar velocities, including those of the supernovae within their hosts, which are negligible at high redshifts and best accounted for at the level of LCs. To account for the motion of nearby objects ($z \lesssim 0.03$), CIGaRS can be seamlessly extended with an appropriate Bayesian hierarchical model (for example, refs. 121–123), which would allow the incorporation of the parameters controlling large-scale structure formation[124,125] within the unified framework.

**Observables of type Ia supernovae.** Observations of supernovae take the form of multi-band LCs: collections of flux measurements in different filters at irregularly spaced times, which vary from supernova to supernova. However, cosmological analyses are typically performed after the LCs have been summarized (independently of one another) by subtracting the constant contribution of the host and fitting an LC model, which produces parameter estimates $\hat{x}(\mathbf{LC})$, $\hat{c}(\mathbf{LC})$ and $\hat{m}(\mathbf{LC})$ and fit uncertainties $\hat{\sigma}_x(\mathbf{LC})$, $\hat{\sigma}_c(\mathbf{LC})$ and $\hat{\sigma}_m(\mathbf{LC})$. It is then assumed that these summary statistics are related to the extrinsic supernova parameters $x_{\rm ext}^s, c_{\rm ext}^s$ and $m^s$ by Gaussian sampling distributions:

$$\hat{x}^s \sim \mathcal{N}\left(x_{\rm ext}^s, (\hat{\sigma}_x^s)^2\right), \quad (20)$$

$$\hat{c}^s \sim \mathcal{N}\left(c_{\rm ext}^s, (\hat{\sigma}_c^s)^2\right), \quad (21)$$

$$\hat{m}^s \sim \mathcal{N}\left(m^s, (\hat{\sigma}_m^s)^2\right). \quad (22)$$

As the uncertainties depend only on instrumental properties, like noise and cadence, their distributions can be robustly determined and treated as fixed simulator inputs. We used the model of Boyd et al.[126], which is based on the results of current surveys and LSST simulations:

$$\ln \hat{\sigma}_x^s \sim \mathcal{N}\left(-1.5, 0.5^2\right), \quad (23)$$

$$\ln \hat{\sigma}_c^s \sim \mathcal{N}\left(-3.5, 0.3^2\right), \quad (24)$$

$$\ln \hat{\sigma}_m^s \sim \mathcal{N}\left(0.1(m^s - 56), 0.6^2\right). \quad (25)$$

We note that these values assume that the LC fits have been performed given a precise redshift measurement from spectroscopy of the supernova or the host. In its absence, for example, when inferring the redshift from the LC itself[127], the fit uncertainties will be inflated, as both colour and (to a lesser extent) stretch are a posteriori correlated with redshift[80] (Fig. 6). This would require adjusting the above distributions accordingly.

It is also possible to extend this description to a dense covariance matrix that represents the systematic correlations between supernovae, as demonstrated in SICRET. However, with hierarchical modelling and SBI, these can, instead, be accounted for explicitly in the simulator, thus replacing the likelihood-centric covariance formulation. Issues related to summary statistics can also be circumvented altogether by extending the simulation and analysis to full LCs, as demonstrated in SIDE-real.

**Detection and selection of type Ia supernovae.** The detection and inclusion of supernovae in modern cosmological analyses is a complex process that considers the raw data quality, the goodness of LC fits and the estimated summaries. Using host observations (for example, to derive redshifts or to study host–type Ia supernova connections) relies on a further procedure that identifies and associates them correctly (for example, refs. 128,129). A big advantage of our framework is that arbitrary detection, selection and association (and classification) criteria can be straightforwardly integrated into the forward model and accounted for with SBI, as recently shown in STAR NRE.

As the present study does not specifically focus on selection effects, we adopt the same simple supernova detection and selection procedure as in STAR NRE (Section 3.2 in ref. 49). Thus, we derive from the expected LSST observing conditions a probability $p(S^s|\hat{m}^s, z^s)$ that depends on the observed brightness of the supernova and its redshift (due to the different detection limits in the different (observer-frame) bands in which the peak occurs). We treat as 'detected and selected' all host–type Ia supernova pairs ('objects') where both $S^h$ and $S^s$ are sampled true. We label their count $N_{\rm sel}$ and treat it as an observable, as in STAR NRE. Finally, the output of the simulator is a set of length-12 vectors that combine the host- and supernova-related observables for each of the detected and selected objects:

$$D \equiv \left\{\left(\mathbf{d}^{h(i)}, \hat{m}^s, (\hat{\sigma}^i_m)^2, \hat{x}^s, (\hat{\sigma}^i_x)^2, \hat{c}^s, (\hat{\sigma}^i_c)^2\right)\right\}_{i=1}^{N_{\rm sel}}. \tag{26}$$

**Implementation details and simulated counts.** Here we discuss two technical details of our forward simulator.

First, we note that Prospector relies on the cosmological model to map redshift to age: $T(z_c, \mathcal{C})$. In principle, this calculation can be repeated for each sampled cosmology in our simulator; however, we do not expect this to have a noticeable effect on our results, as type Ia supernovae are mainly informative of cosmological distances rather than times; that is, for values of $\mathcal{C}$ consistent with a given sample of type Ia supernovae, $T(z_c, \mathcal{C})$ does not vary significantly with $\mathcal{C}$. Therefore, we treat $T(z_c)$ as fixed to that consistent with the cosmology from the Wilkinson Microwave Anisotropy Probe[130], as when training and creating Prospector-β.

The second point concerns the number of objects that we simulate, which, in principle, depends on (and, hence, is informative of) the astrophysical and cosmological models (for example, through a parameter $S_8$), as well as the surveyed sky area ($\Omega$). At present, our high-level supernova-focused simulator does not include these details, as supernova observations do not carry information about this connection, beyond the distribution of their redshifts, which we assume is extracted mainly from the hosts.

Hence, the purely host-related part of CIGaRS has no free global parameters up to the point where the redshifted (observer-frame) spectral flux

$$\Phi^h_{\rm o}(\lambda_{\rm o}) \equiv \frac{\Phi^h(\lambda_{\rm o}/(1+z^h))}{(1+z^h)^3} \tag{27}$$

is converted to the spectral flux density through cosmological distance, as in equation (4). This means that we can generate a fixed 'bank' of galaxies and associated observables—noiseless absolute ugrizy magnitudes obtained by integrating equation (27) through the respective band-pass filter—to serve as potential hosts when simulating supernovae. We choose a bank size of 1,000,000 and note that this represents the total population of galaxies ($N_{\rm gal}$).

However, as we already noted in Fig. 3, the distribution of transient hosts is significantly skewed towards high-mass galaxies due to the abundance of prospective progenitors within them. As these are a minority in the total population and we use a fixed simulation bank, we, thus, run the risk of repeatedly including the same objects in our training catalogues, which could cause the network to become overfitted. To prevent this, we modified the distribution from which we sampled galaxy properties when generating the simulation bank to more closely represent the final distribution of type Ia supernova hosts. Specifically, we modified the stellar mass function in Prospector-β:

$$p(M_*, z_c) \to \frac{M_*}{1+z_c} p(M_*, z_c), \tag{28}$$

motivated by equation (8). We then exactly undid this when calculating the number of expected supernovae within each host by dividing equation (8) by the same factor, which preserves $\langle N_{\rm SN}\rangle$ as a function of mass and redshift.

Then, when compiling a mock survey, we calculated the apparent brightnesses of the galaxies (given $\mathcal{C}$), added noise and evaluated detection as in section 'Photometry and detection of hosts'. We then seeded only the detected and selected subset with type Ia supernovae as described in section 'Occurrence of type Ia supernovae'. This is an implicit selection criterion on the supernovae: we only considered supernovae for which we can observe and uniquely identify the host (with certainty); for our (current) method to be applicable, we need to apply the same cut to the real data, but this is usually not a stronger requirement than the cuts typically applied on the supernova data. For the default cosmological model and DTD (Extended Data Table 1), only about 39% of the galaxies are selected, and of these, the type Ia supernova rate is ~2,500 yr$^{-1}$, of which about 1/3 pass detection and selection (for the default population parameters and our toy model of LSST).

Last, we need to set a survey duration ($\mathcal{T}$ equation (8)), but this is only well defined in combination with $\Omega$ or a physically meaningful $N_{\rm gal}$, which we replaced with the fixed size of the galaxy 'bank'. Therefore, we set an arbitrary scaling in equation (8) so as to achieve (on expectation) a given number of detected and selected objects, for example, 1,600 or 16,000. Importantly, as $\Omega$ and $\mathcal{T}$ are well known for real surveys, we can easily extend the simulator to properly calculate the galaxy and type Ia supernova counts when analysing real data. In the present set-up, we can apply the same scaling when generating training data as for the test mocks.

## Set-based TMNRE for hierarchical models

Bayesian hierarchical models feature a large number of free parameters, which scales with the number of objects observed. Accurate, precise and fast inference from future data is, thus, a considerable computational challenge that likelihood-based methods are ill-suited to address. Moreover, the likelihood requires explicit calculations of intractable probabilities, often leading to ad hoc approximations. SBI addresses all these issues—scalability, realism and rigour—by delivering marginal Bayesian posteriors, given data $D_0$, for any parameters of interest $\boldsymbol{\theta}$, implicitly integrated over all nuisance parameters ($\boldsymbol{\nu}$) and relevant stochastic processes (for example, selection and other systematic effects):

$$p(\boldsymbol{\theta}|\mathbf{D}_0) \propto p(\boldsymbol{\theta})p(\mathbf{D}_0|\boldsymbol{\theta}) = p(\boldsymbol{\theta}) \int p(\mathbf{D}_0|\boldsymbol{\nu},\boldsymbol{\theta})\, p(\boldsymbol{\nu}|\boldsymbol{\theta})\, \mathrm{d}\boldsymbol{\nu}. \tag{29}$$

SBI requires only samples from the prior ($p(\boldsymbol{\theta})$) and the marginal likelihood ($p(\mathbf{D}|\boldsymbol{\theta})$), which are provided by a stochastic forward simulator that represents the Bayesian joint model $p(\boldsymbol{\theta}, \mathbf{D})$. There are different techniques (flavours of neural SBI) for using simulated pairs $(\boldsymbol{\theta}, \mathbf{D})$ to train a NN and later perform inference from observed data $\mathbf{D}_0$. We adopt the approach called neural ratio estimation[74], as it offers the greatest

freedom in the choice of the NN architecture and in choosing priors for training and evaluation. It trains a network $\hat{r}(\boldsymbol{\theta}, \mathbf{D})$ to approximate the single real number

$$r(\boldsymbol{\theta}, \mathbf{D}) \equiv \frac{p(\boldsymbol{\theta}, \mathbf{D})}{p(\boldsymbol{\theta})p(\mathbf{D})} = \frac{p(\boldsymbol{\theta}|\mathbf{D})}{p(\boldsymbol{\theta})} \tag{30}$$

by minimizing the binary cross-entropy loss commonly used for classification tasks. Once trained, $\hat{r}(\boldsymbol{\theta}, \mathbf{D}_0)$ evaluated with the observed data can simply be multiplied by the prior or used to reweight prior samples to represent the target posterior, according to the second equality in equation (30).

In principle, $\boldsymbol{\theta}$ can represent any group of parameters that we wish to derive a posterior for. However, scientific interpretations are usually based on one- or two-dimensional marginal distributions, and consequently, we will define the following groups of (global) parameters of interest $\gamma_g$:

- one-dimensional: $M_0, \sigma_0, \alpha, \beta, \alpha_c, M_*^{\text{step}}$;
- two-dimensional: $[A, b]$ for the DTD and $[\Omega_{m0}, \Omega_{\Lambda 0}]$ for cosmology;
- two-dimensional: $[\Delta M, \gamma_{Z_*}]$, $[\Delta M, \gamma_{t_*}]$, and $[\gamma_{Z_*}, \gamma_{t_*}]$, forming all combinations of the host–type Ia supernova connection parameters: a so-called corner plot.

For each of them, we train—simultaneously—separate ratio estimators $\hat{r}_g$, following shared data preprocessing.

**Object-specific parameters.** In addition to the global parameters, we will—simultaneously—train ratio estimators for all $\mathcal{O}(N_{\text{sel}})$ object-specific (local) parameters in the hierarchical model (here we will use the unified label $\boldsymbol{\lambda}^i$, which represents $\mathbf{g}^h$ and $\boldsymbol{\lambda}^s$ in CIGaRS collectively). Although we can, as before, form local-parameter groups $\lambda_g$ from parameters for the same object, we will infer each object-specific parameter $\lambda_g \in \{z_c, M_*, Z_*, \delta, \tau_2, x_{\text{int}}, c_{\text{int}}\}$ marginally. Although in Fig. 6 we show only a subset of the results that represent scientific interest in the present study, estimating all local parameters during training helps extract informative features and ultimately improves the inference of the global parameters.

In SICRET and SIDE-real, we demonstrated simultaneous marginal inference of all $\{\boldsymbol{\lambda}^i\}$ for models in which the dataset size $N_{\text{sel}}$ was known a priori. That is, it could be fixed after a survey is performed because selection effects were not considered. Here we extend the approach to simulators that produce datasets with various $N_{\text{sel}}$, which introduces two complications.

The first concerns identifying the objects. Within Bayesian hierarchical modelling, the $\{\lambda^i\}$ are a priori independent and identically distributed, conditional on the set of global parameters $\boldsymbol{\gamma}$, and each influences the sampling distribution of only one observed 'object'; that is we have the general model structure:

$$p(\gamma, \{\lambda^i\}, \{\mathbf{d}^i\}) = \left[\prod_{i=1}^{N_{\text{sel}}} p(\mathbf{d}^i|\lambda^i, \boldsymbol{\gamma})p(\lambda^i|\boldsymbol{\gamma})\right] p(\gamma). \tag{31}$$

Previously, we used a fixed ordered collection of auxiliary variables $[\mathbf{a}^i]_{i=1}^{N_{\text{sel}}}$ to disambiguate the assignment of labels $i$, which effectively individualized the sampling distributions:

$$p(\mathbf{d}^i|\lambda^i, \boldsymbol{\gamma}) \to p(\mathbf{d}^i|\mathbf{a}^i, \lambda^i, \boldsymbol{\gamma}) \to p_i(\mathbf{d}^i|\lambda^i, \boldsymbol{\gamma}). \tag{32}$$

However, the present simulator produces unordered (exchangeable) sets of a priori undetermined sizes, and so the auxiliary variables need to become an output of the model, that is be incorporated into the observable $\mathbf{d}^i$. Indeed, CIGaRS explicitly models the host photometry (from which the redshift is ultimately derived) and the observational (fitted) uncertainties, which comprised $\mathbf{a}^i$ in SICRET and SIDE-real. Thus, we can treat $\boldsymbol{\lambda}^i$ and $\mathbf{d}^i$ as realizations of singular random variables $\boldsymbol{\lambda}$ and $\mathbf{d}$:

$$\left[\prod_i p(\mathbf{d} = \mathbf{d}^i|\boldsymbol{\lambda} = \lambda^i, \boldsymbol{\gamma})p(\boldsymbol{\lambda} = \lambda^i|\boldsymbol{\gamma})\right] p(\boldsymbol{\gamma}), \tag{33}$$

rather than collections of separate (independent and identically distributed) random variables, which allows us to train a single network to represent all posteriors:

$$\mathrm{p}(\lambda^i \,|\, \{\mathbf{d}^i\}) \to \mathrm{p}\left(\lambda \,\middle|\, \mathbf{d} = \mathbf{d}^i, \cancel{\{\mathbf{d}^i\}}\right), \tag{34}$$

where we have ignored the dependence on the full $\{\mathbf{d}^i\}$, as it is approximately redundant with the truncation of global parameters, as argued for SICRET.

Second, a technical complication arises as we need training pairs

$$\boldsymbol{\lambda}, \mathbf{d} \sim p(\boldsymbol{\lambda}, \mathbf{d}) = \int p(\boldsymbol{\lambda}, \mathbf{d}|\boldsymbol{\gamma})p(\boldsymbol{\gamma})\,\mathrm{d}\boldsymbol{\gamma}, \tag{35}$$

but the simulator samples in proportion to $\langle N_{\text{sel}}\rangle(\boldsymbol{\gamma}) \times p(\boldsymbol{\gamma})$ rather than simply $p(\boldsymbol{\gamma})$. This is not a problem when $N_{\text{sel}}$ is a fixed input and can otherwise be rectified by randomly selecting a fixed number $N_{\text{sub}}$ of objects from the simulator output. In fact, $N_{\text{sub}} := 1$ would be the natural choice in line with the above treatment of $\{\boldsymbol{\lambda}^i\}$ and $\{\mathbf{d}^i\}$ as realizations of $\boldsymbol{\lambda}$ and $\mathbf{d}$, and we chose this when generating a validation set, which also represents the prior $p(\boldsymbol{\lambda})$. Still, by setting a larger $N_{\text{sub}}$, we can cheaply (without extra simulation) enlarge the effective batch size for training local-parameter inference networks, and so we chose $N_{\text{sub}} = 100$ in this case. Finally, when evaluating the results for the $N_{\text{sel},0}$ objects in $\boldsymbol{D}_0$, we simply skipped the subselection and used the full set ($N_{\text{sub}} := N_{\text{sel},0}$).

**Network architecture: conditioned deep set++.** The NN we use (depicted, together with details of the sizes of its various layers, in Fig. 4) is based on the conditioned deep-set architecture from STAR NRE (following Zaheer et al.[76]) but augmented in depth as in Zhang et al.[75] due to the sophistication of the simulator and the inference tasks and with the addition of local-parameter estimators as in SICRET and SIDE-real.

Given a dataset $\boldsymbol{D} \equiv \{\mathbf{d}^i\}$ with cardinality $N_{\text{sel}}$, we first preprocess each element $\mathbf{d}^i$ with a small feed-forward network to (automatically) derive (nonlinear) 'features':

$$\mathrm{d}^i = \mathtt{DataPre}(\mathbf{d}^i), \tag{36}$$

which are generally useful across all inference tasks. These may include galaxy colours, (fiducially) standardized supernova magnitudes, observational uncertainty (from the provided variances) and estimates of object-specific parameters directly usable in the respective downstream ratio estimators. Although they are not forced to be directly 'meaningful' to a human scientist, we plan to explore and interpret these features in future work. The $\{\mathrm{d}^i\}$ are then used as the input for all ratio estimators.

Following Zaheer et al.'s representation theorem[76] and the considerations laid out in STAR NRE, the estimator for a group of global parameters $\boldsymbol{\theta}_g$ makes use of two feed-forward networks $\hat{\phi}_g$ and $\hat{\rho}_g$ and takes the form:

$$\ln \hat{r}_g(\boldsymbol{\theta}_g, \mathbf{D}) \equiv \hat{\rho}_g(\boldsymbol{\theta}_g, \mathrm{S}_g, N_{\text{sel}}), \tag{37}$$

with

$$\mathrm{S}_g \equiv \frac{1}{N_{\text{sel}}} \sum_i \hat{\phi}_g\left(\boldsymbol{\theta}_g, \frac{\mathbf{d}^i - \mathtt{m}}{\mathtt{s}}, \mathtt{m}, \mathtt{s}\right), \tag{38}$$

where $\mathtt{m}$ and $\mathtt{s}$ are, respectively, the mean and standard deviation of the $\{\mathrm{d}^i\}$. This implements the `SetNorm` operation advocated by Zhang et al.[75] and explicitly preserves the information contained in the set moments $\mathtt{m}$ and $\mathtt{s}$. For expressivity (which allows the set aggregation to be a simple averaging operation), $\hat{\varphi}_g$ is a deep residual network that implements 'skip connections' as proposed by Zhang et al.[75] after combining its inputs ($\boldsymbol{\theta}_g$, $\frac{\mathbf{d}^i - \mathtt{m}}{\mathtt{s}}$, $\mathtt{m}$, $\mathtt{s}$) into one vector through a single hidden network layer (Fig. 4). Finally, $\hat{\rho}_g$ similarly combines its inputs ($\boldsymbol{\theta}_g$, $\mathtt{S}_g$, $N_{sel}$) and feeds them through a few fully connected layers to output the single number $\ln \hat{r}_g(\boldsymbol{\theta}_g, \mathbf{D})$.

The ratio estimator for a local or object-specific parameter $\lambda_g$ is simpler, as it does not need to aggregate information across the set, and so it consists of a single feed-forward network:

$$\ln \hat{r}_g(\lambda_g, \mathbf{d}^i) = \mathtt{LocalNRE}_g(\lambda_g, \mathbf{d}^i), \tag{39}$$

which can be applied in parallel over the $N_{sub}$ input data at the same time.

**Truncation and fine-tuning.** Neural SBI is amortized: once the network is trained, results can be quickly derived from numerous simulated (or real, if available) data realizations. While this can be exploited to verify and calibrate inference (see, for example, refs. 57,58), that would require training over a wide range of data, which can be inefficient if one is interested only in the results for a single dataset $\boldsymbol{D}_0$.

In such cases, the simulation budget, training time and network capacity can be optimized through various sequential SBI strategies, which modify (either continuously or in a succession of stages) the prior $p(\boldsymbol{\theta}) \to \tilde{p}(\boldsymbol{\theta})$ from which parameters are drawn, based on intermediate results during training. A simple yet effective prescription is prior truncation[73], in which the shape or form of $\tilde{p}(\boldsymbol{\theta})$ is unchanged, but its support is restricted to a region $T(D_0)$ in which the posterior density is significantly different from zero:

$$\boldsymbol{\theta} \notin T(D_0) \;\Rightarrow\; p(\boldsymbol{\theta}|D_0) \approx 0, \tag{40}$$

as approximated by a previously trained network evaluated at $D_0$. Thus, the only modification needed is a trivial renormalization to account for the excluded prior probability mass:

$$p(\boldsymbol{\theta}) \to \tilde{p}_{T(\mathbf{D}_0)}(\boldsymbol{\theta}) = \begin{cases} \frac{p(\boldsymbol{\theta})}{\int_{T(\mathbf{D}_0)} p(\boldsymbol{\theta})\,\mathrm{d}\boldsymbol{\theta}} & \text{if } \boldsymbol{\theta} \in T(\mathbf{D}_0), \\ 0, & \text{otherwise,} \end{cases} \tag{41}$$

which leaves the posterior (given $\boldsymbol{D}_0$) unchanged while restricting simulated examples to more closely resemble $\boldsymbol{D}_0$.

We apply truncation separately for each global inference group. In one dimension, we use a simple contiguous interval that contains 99.99% approximate credibility and then sample training examples by analytically modifying the (simple, whether uniform or normal) prior. For two-dimensional groups, we either define an iso-likelihood contour (which again contains 99.99% of approximate posterior mass) and use rejection sampling within it or apply one-dimensional truncation separately for the two parameters (we truncate to an axis-aligned rectangular region) when they are not strongly correlated. The truncation regions are illustrated with grey shading in Fig. 5.

Once new training data are generated, we resume—instead of re-initializing—training, which means that the network is simply fine-tuned to give more accurate results in the 'zoomed-in' parameter space, instead of being forced to relearn the analysis from scratch. This is especially relevant for the preprocessor and deep-set featurizers (described below), whose computations should not appreciably depend on the parameter ranges.

**Training details.** Analysing the mock data with ~1,600 objects required two rounds of truncation (starting from the priors in Extended Data Table 1 and with the network in Fig. 4) for the results to converge (judged by the difference between posteriors from successive rounds). In each round, we generated 64,000 full example datasets (with 6,400 more for validation and to represent the priors when evaluating results) and trained using the Adam optimizer[131] with learning rate $10^{-3}$ (reduced to $10^{-4}$ when fine-tuning) for at most 200 epochs, although the loss typically plateaus earlier. On two NVIDIA A100 GPUs with a combined mini-batch size of 64, one stage takes ~24 h, with most of the computation taken up by backpropagating through all 11 deep sets (for the separate global-parameter groups).

For the larger ($N_{sel} \approx 16{,}000$) dataset, we took the final network from above and fine-tuned it on simulations from the same final truncated prior region but with a tenfold increase in output size (on expectation). As the deep-set architecture processes the full dataset at once, training it now required roughly ten times more compute and memory, so we used eight GPUs and a batch size of 32, which again required 24 h per truncation stage (we needed only one in this case).

Finally, for each global-parameter group $\boldsymbol{\gamma}_g$, we picked the checkpoint with the lowest validation loss for the specific group, whereas for object-specific parameters $\lambda_g$, we picked the best checkpoint overall (lowest loss averaged over all groups). To represent the posteriors, we evaluated $\hat{r}(\boldsymbol{\theta}_g, \mathbf{D}_0)$ over the parameter samples in the validation set, which represent $p(\boldsymbol{\theta}_g)$, and used the results as importance weights for plotting contours and calculating posterior moments.

## Data availability

Scripts with the implementation of CIGaRS, trained networks, mock data and results are available via GitHub at https://github.com/snai-analysis/cigars and are available via Zenodo at https://doi.org/10.5281/zenodo.18705764 (ref. 132).

## Code availability

CIGaRS is based on the Clipppy convenience layer for inference and probabilistic programming in Python (https://github.com/kosiokarchev/clipppy). The forward simulator is implemented in Pyro[133] and makes use of the cosmographic utilities of phytorch (https://github.com/kosiokarchev/phytorch), the prospector code (https://github.com/bd-j/prospector) and the pretrained Speculator-α emulator (https://github.com/justinalsing/speculator), embedded in SLICsim (https://github.com/kosiokarchev/slicsim). The NNs were defined and trained using PyTorch[134] and PyTorch Lightning[135].

## Acknowledgements

R.T. acknowledges funding from Next Generation EU through the National Recovery and Resilience Plan, Investment PE1 under Project FAIR (Future Artificial Intelligence Research), which is co-financed by the Next Generation EU (Grant No. DM 1555/11.10.22). R.T. is partially supported by the Fondazione ICSC, Spoke 3 'Astrophysics and Cosmos Observations' through the National Recovery and Resilience Plan under Project ID CN00000013, 'Italian Research Center on High-Performance Computing, Big Data and Quantum Computing', which is funded by MUR Missione 4 Componente 2 Investimento 1.4: Potenziamento strutture di ricerca e creazione di 'campioni nazionali di R&S (M4C2-19)' through Next Generation EU. Funding for the work of R.J. was partially provided by Project PID2022-141125NB-I00

through Grant No. CEX2024-001451-M, which is funded by MICIU/AEI/10.13039/501100011033. This work was supported by a grant from the Simons Foundation (Grant Award ID BD-Targeted-00017375, UB).

## Author contributions

K.K. led the research and its implementation with guidance from R.T. and R.J. All authors contributed to producing the paper.

## Funding

Open access funding provided by Scuola Internazionale Superiore di Studi Avanzati - SISSA within the CRUI-CARE Agreement.

## Competing interests

The authors declare no competing interests.

## Additional information

**Extended data** is available for this paper at https://doi.org/10.1038/s41550-026-02842-5.

**Supplementary information** The online version contains supplementary material available at https://doi.org/10.1038/s41550-026-02842-5.

**Correspondence and requests for materials** should be addressed to Konstantin Karchev.

**Peer review information** *Nature Astronomy* thanks the anonymous reviewers for their contribution to the peer review of this work. Peer reviewer reports are available.

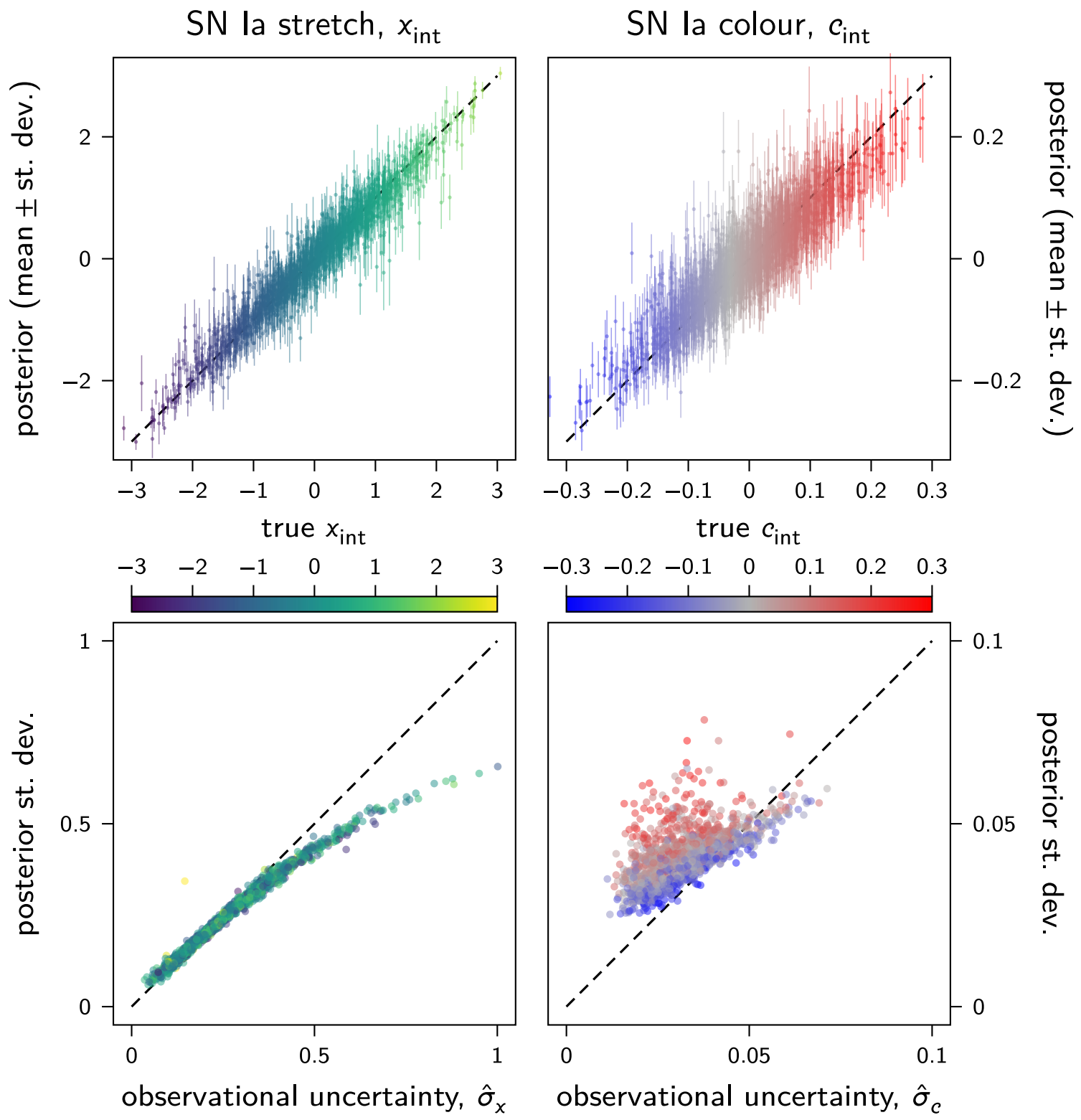

**Extended Data Fig. 1 | Inference results for (true/intrinsic) stretch and colour of the 1578 SNe Ia in our mock data.** Top: Marginal posterior means and standard deviations vs. the true values in the simulation, as in Fig. 6. Bottom: the posterior standard deviations (sizes of the error bars in the top plots) vs. the observational uncertainties (for example from a light-curve fit), which we treat as part of the analysed data. The colour scheme corresponds to the true parameter values (the abscissa in the top plots). Note the effect of shrinkage for objects with large observational uncertainty: the posteriors are more strongly constrained (smaller st. dev.) than the scatter in the data ($\hat{\sigma}$) due to the known population distribution. Note further that inference of SN colour is complicated by the presence of dust, which manifests as an increase in posterior uncertainty, and that this effect is larger for intrinsically redder objects, for which the colour can be attributed to dust reddening.

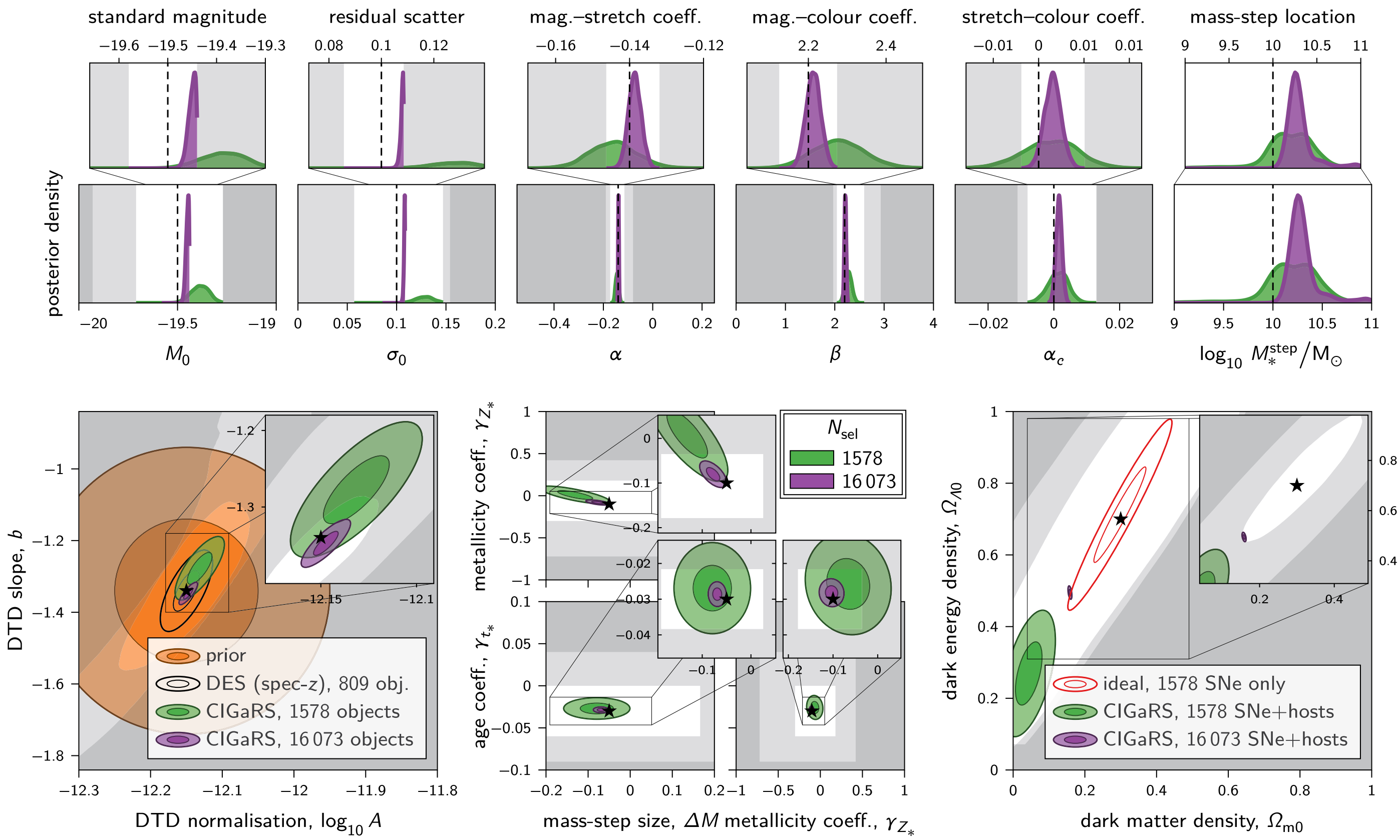

**Extended Data Fig. 2 | Inference with a mis-specified model.** Marginal posterior densities and regions with 39 % and 86 % marginal posterior credibilities, respectively, for one and two-dimensional global-parameter groups (as in Fig. 5), derived with the same networks used in the main text but from mock observations generated with the F99[82] dust law instead of KC13[81], which was used for training. The resulting bias is most pronounced for parameters affecting the Hubble diagram ($M_0$, $\sigma_0$, and cosmology) and increases when the data set is larger and more constraining. Note that we have not retrained the model using constrained priors appropriate for the new data, which results in some posteriors ($M_0$, $\sigma_0$, and cosmology) appearing at the edges of the respective (truncated) prior regions: one of the signs that the model may be mis-specified. With SBI, one can include the systematic uncertainty from modelling choices through model averaging during training, that is by simulating with a diverse set of dust laws, etc.

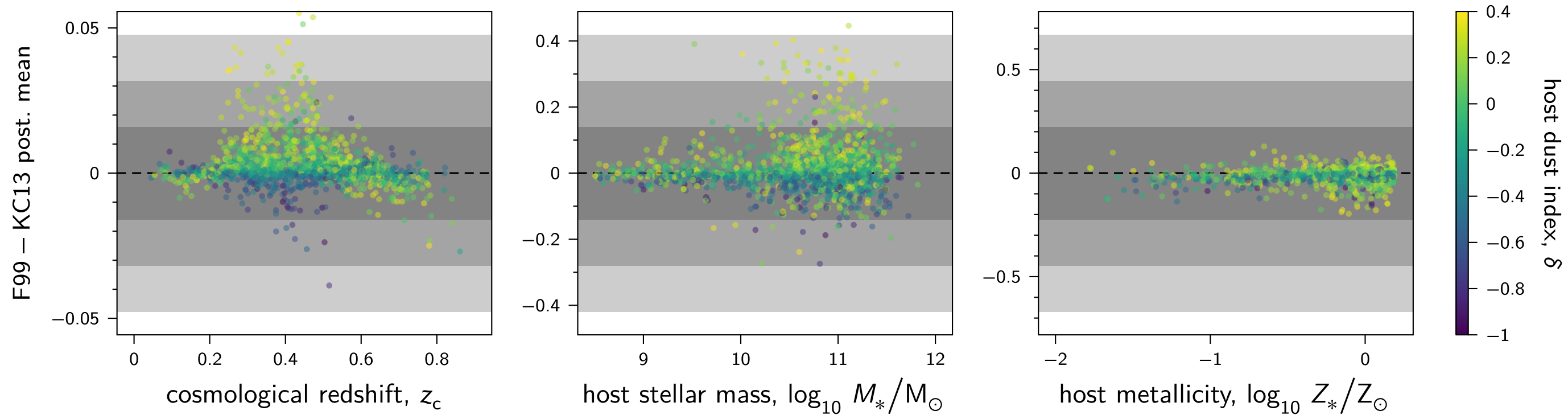

**Extended Data Fig. 3 | Change in the results for object-specific parameters.** Shift of the marginal posterior means plotted against the true parameter values and coloured by true dust-law parameter $\delta$ when mock data is generated with the F99[82] dust law but analysed with that of KC13[81] (as in Extended Data Fig. 2). Changes are only mildly significant at the individual-object level: compare with the scatter from the true values, which we show as 1-, 2-, and 3-$\sigma$ shaded areas as in the middle panels of fig. 6 (compare also with the uncertainties in the bottom panels there), but being correlated, they can dramatically affect the inference of global parameters.

**Extended Data Table 1 | Parameters in the model, their (hierarchical) priors, and the values used to simulate the analysed mock data**

| | parameter | symbol | value / range | unit | prior |
|---|---|---|---|---|---|
| $\mathcal{C}$ | DM density ($z_c = 0$) | $\Omega_{m0}$ | 0.3 | — | $\mathcal{U}(0;1)$ |
| | DE density ($z_c = 0$) | $\Omega_{\Lambda 0}$ | 0.7 | — | $\mathcal{U}(0;1)$ |
| galaxy model | cosmological redshift | $z_c^h$ | [0.02; 1.52] | — | Prospector-β |
| | stellar mass (total) | $\log_{10} M_*^h/M_\odot$ | [8.5; 12.5] | dex | Prospector-β |
| | stellar metallicity | $\log_{10} Z_*^h/Z_\odot$ | [−1.98; 0.19] | dex | Prospector-β |
| | dust optical depth | $A_V^h$ | [0; 4] | mag | eqn.(2) |
| | KC13 dust law | $\delta^h$ | [−1; 0.4] | — | eqn.(3) |
| DTD | normalisation | $\log_{10} A$ | −12.15 | dex | $\mathcal{N}(-12.15, 0.1^2)$ |
| | slope | $b$ | −1.34 | — | $\mathcal{N}(-1.34, 0.2^2)$ |
| "causual" host–SN Ia connection | "mass step" size | $\Delta M$ | −0.05 | mag | $\mathcal{U}(-0.2;0.2)$ |
| | "mass step" location | $\log_{10} M_*^{step}/M_\odot$ | 10 | dex | $\mathcal{U}(9;11)$ |
| | magnitude–metallicity | $\gamma_{Z_*}$ | −0.1 | mag dex$^{-1}$ | $\mathcal{U}(-2;2)$ |
| | magnitude–age | $\gamma_{t_*}$ | −0.03 | mag Gyr$^{-1}$ | $\mathcal{U}(-0.1;0.1)$ |
| | "causal" mag.offset | $\delta M^s$ | $(-\infty;\infty)$ | mag | eqn.(10) |
| SN Ia pop. | abs.magnitude ($B$-band) | $M_{int}^s$ | $(-\infty;\infty)$ | mag | $\mathcal{N}(M_0 + \alpha x_{int}^s + \beta c_{int}^s + \delta M^s, \sigma_0^2)$ |
| | standard abs.mag. | $M_0$ | −19.5 | mag | $\mathcal{U}(-20;-19)$ |
| | residual mag.scatter | $\sigma_0$ | 0.1 | mag | $\mathcal{U}(0.01;0.2)$ |
| | "stretch" | $x_{int}^s$ | $(-\infty;\infty)$ | — | $\mathcal{N}(\mu_x, \sigma_x^2)$ |
| | pop.mean | $\mu_x$ | 0 | — | fixed |
| | pop.st.dev. | $\sigma_x$ | 1 | — | fixed |
| | "colour" | $c_{int}^s$ | $(-\infty;\infty)$ | mag | $\mathcal{N}(\mu_c + \alpha_c x_{int}^s, \sigma_c^2)$ |
| | pop.mean | $\mu_c$ | 0 | mag | fixed |
| | pop.st.dev. | $\sigma_c$ | 0.1 | mag | fixed |
| intrinsic correlation coefficients | magnitude–stretch | $\alpha$ | −0.14 | mag | $\mathcal{U}(-0.55;0.25)$ |
| | magnitude–colour | $\beta$ | 2.2 | — | $\mathcal{U}(0;4)$ |
| | colour–stretch | $\alpha_c$ | 0 | mag | $\mathcal{U}(-0.03;0.03)$ |
| extrinsic effects | extinction ($V$-band) | $A_V^s$ | $[0;+\infty)$ | mag | Exponential($1/\tau$) |
| | pop.mean | $\tau$ | 0.1 | mag | fixed |
| | total-to-selective extinction | $R_V^s$ | $[1.2;+\infty)$ | — | $\mathcal{N}(\mu_R, \sigma_R^2)$ |
| | pop.mean | $\mu_R$ | 3 | — | fixed |
| | pop.st.dev. | $\sigma_R$ | 0.5 | — | fixed |

The given ranges indicate the support of the respective priors, within which object-specific parameters are drawn.